\documentclass[a4paper,11pt]{article}
\input epsf
\usepackage{graphics}
\begin{document}
\renewcommand{\arraystretch}{0.666666666}
%\parindent=0pt
%{\large
\parskip.2in
\newcommand{\be}{\begin{equation}}
\newcommand{\ee}{\end{equation}}
\newcommand{\acc}{\\[3mm]}
\newcommand{\cmod}[1]{ \vert #1 \vert ^2 }
\newcommand{\ie}{{\em ie }}
\nopagebreak[3]

\title{ Solitons in $\alpha$-helical proteins\\
}
\author{
L. Brizhik\thanks{e-mail address: brizhik@bitp.kiev.ua},\,
and
A. Eremko\thanks{e-mail address: eremko@bitp.kiev.ua},\,
\\
Bogolyubov Institute for Theoretical Physics, 03143 Kyiv, Ukraine
\acc
\acc
B. Piette\thanks{e-mail address: B.M.A.G.Piette@durham.ac.uk} \,
and
W. Zakrzewski\thanks{e-mail address: W.J.Zakrzewski@durham.ac.uk}
\\
Department of Mathematical Sciences,University of Durham, \\
Durham DH1 3LE, UK\\
}
\date{}

\maketitle

\begin{abstract}
We investigate some aspects of the soliton dynamics
 in an $\alpha$-helical protein 
macromolecule within the steric Davydov-Scott model. Our main 
objective is to elucidate the important role of the helical symmetry in 
the formation, stability and dynamical properties of Davydov's solitons in 
an $\alpha$-helix. We show, analytically and numerically, that 
the corresponding system of nonlinear equations admits several types of 
stationary soliton solutions and that solitons which 
preserve helical symmetry are dynamically unstable: once  
formed, they decay rapidly when they propagate. 
On the other hand, the soliton which spontaneously 
breaks  the local translational and helical 
symmetries possess the lowest energy and is a robust localized entity.
 We also demonstrate that this soliton is the result of an hybridization of 
the quasiparticle states from the two lowest degenerate bands 
and has an inner structure which can be described as a modulated multi-hump 
amplitude distribution of excitations on individual spines. The 
complex and composite structure of the soliton manifests itself 
distinctly when the soliton is moving  and some interspine 
oscillations take place. Such a soliton structure and the interspine oscillations 
have previosly been observed numerically in [A.C. Scott, Phys.Rev. A {\bf 
26} 578 (1982)]. Here we argue that the solitons studied by A. Scott, 
are hybrid solitons and that the oscillations arise 
due to the helical symmetry of the system and result from the motion 
of the soliton along the $\alpha$-helix. The frequency of the interspine 
oscillations is shown to be proportional to the soliton velocity.

\end{abstract}
\medskip

\section{Introduction}

In the 1970s Davydov \cite{ASD73}    
proposed a nonlinear mechanism for the storage and transfer of   
vibrational energy (intrapeptide vibration Amid-I) in    
alpha-helical proteins. As a result of the interaction of   
high-frequency Amid-I vibrations (vibrations of double C-O bond of   
peptide groups) with the low-frequency acoustic vibrations of the   
protein, a self-trapping of the Amid-I vibration takes place.   
This idea has attracted a lot of interest,   
which has increased even further after the appearance of a paper 
\cite{Dav-Kisl1} in which Davydov and Kislukha   
demonstrated that the corresponding system of equations for a   
molecular chain admits, in the continuum approximation, a solitonic   
solution.  This solution describes a self-trapped quasiparticle  
(a lump of vibrational Amid-I energy) that propagates at constant velocity  and 
is acompanied by a self-consistent chain deformation \cite{Dav-Kisl2}.

Since then, various properties of such one-dimensional polaron-like 
self-trapped states have been studied in detail both analytically and 
numerically (see, e.g., \cite{ASD,nato,ASc}).  Dynamical properties 
of Davydov solitons and their formation, given various initial 
conditions of the chain, have been investigated in discrete chains 
and in continuum models. Most of these results have been obtained for 
a single chain. Often they have involved numerical values of the 
parameters that are characteristic of real proteins; thus very often 
 these results have been discussed in the context of an 
$\alpha$-helix.  In reality, however, real $\alpha-$helices contain 
three strands, each of which, contains periodically placed peptide 
groups connected by hydrogen bonds. A three-strand model for an $\alpha 
$-helix was proposed in \cite{DavSoopr}, where the stationary states 
were first studied. This model represents an $\alpha $-helix as a 
three-strand structure with three peptide groups per cell in a plane 
perpendicular to the protein axis. Soon afterwards the properties of 
such soliton states were studied analytically in \cite{DES,Er-Serg} 
and numerically in \cite{Hyman,Hen}. 

This model does not include a helical structure of proteins, and so
  afterwards the model was improved in \cite{Sct1,FedYak}.  In 
\cite{FedYak} the soliton solutions which do not  break the 
chiral symmetry were found analytically.  So far, the most complete 
numerical study of the problem has been presented by Scott in 
\cite{Sct1}, where the formation of a soliton in a linear chain of a 
finite length had been investigated using the initial excitation of a 
certain form localised on two of the three peptide groups at the end 
of the chain.  Scott showed there that, under such conditions, a 
soliton can be formed and that this soliton propagates along the 
protein with a constant velocity.  It has turned out that such a 
soliton has an inner structure and that some interchain oscillations 
of energy take place.  These oscillations were compared by Scott with 
the lines in experimentally measured Raman spectra of living cells 
(see also \cite{Sct2}). Let us add here also that in this numerical 
modelling only one type of initial excitation was used, and, as a 
result, only one value of the velocity of the soliton of a given 
symmetry was obtained. However,  there are several features which 
suggest that this picture is oversimplified. In fact, we expect the 
dynamics to contain some oscillatory  features.  This is due to the 
 discrete nature of the chain and it  can also be related to the 
helical symmetry of the protein and the symmetry of the initial 
excitation. 

The aim of the present paper is to study the soliton 
states in an $\alpha$-helix, to investigate their properties and their
stability and  to look at the dependence of the internal soliton  
vibrations on the velocity of the soliton propagation. In Section 2 a 
general description of the model is given. In Section 3 the 
elementary excitations of the $\alpha-$helix are presented. In 
Section 4 we describe the results of our analytical studies of  
soliton states in the adiabatic approximation while the results of 
our numerical modelling are presented in Section 5. In Section 6 we discuss 
the applicability of the adiabatic approximation and in Conclusions we make further comments on the physical relevance of our results.

\section{The general model}

Protein macromolecules are long nonbranched polymer chains which are
formed as a result of polymerization of aminoacids. Aminoacid
residues in such polymer chains are connected by the {\it peptide bonds} in
which four atoms (OCNH) form the {\it peptide group} and two
$\alpha$-carbon atoms of the residues are placed in one plane. So the
backbone of such a polypeptide chain can be described as a set of comparably
rigid planes divided by methilene groups (-CHR-). Because peptide
groups (PGs) are bonded with methilene groups by ordinary bonds, a free
rotation of PG planes around these bonds is possible. Due to such
rotations, a polypeptide chain can take different spatial
configurations. Thus in particular, it can be rolled into a helix. Such a
configuration of the polypeptide chain is stabilized by the intrachain hydrogen
bonds which are formed between a hydrogen atom of a PG and an oxygen atom
of the fourth group along the chain. Such a helical structure,
called $\alpha$-helix, has 3.6 peptide groups per turn. Thus, the equilibrium 
positions of the repeated units (PG) in an $\alpha$-helix are determined by the
radius-vectors 
\begin{equation}
\vec{R}_{l}^{(0)} = r \left( \vec{e}_x \cos(\frac{2\pi l}{3.6}) + 
\vec{e}_y \sin(\frac{2\pi l}{3.6})\right) + \vec{e}_z \frac{a l}{3.6} 
\label{aleqvpos}
\end{equation}
where $\vec{e}_i$ ($i = x,y,z$) are unit vectors along coordinate 
axes, $a$ is a period of the helix, $r$ is its radius, and $l$ is an integer
labeling each group along the polypeptide chain.  
The nearest 
neighbours (sites $l$ and $l \pm 1$) along the chain are bound by  
rigid valence bonds and each $l$-th group in a helix is bound with $(l 
\pm 3)$-th groups by soft hydrogen bonds forming three spines along 
the helix.

The three spines along the $\alpha$-helix are formed by units with numbers:
\be
l_1 = 3n - 1,\quad  l_2 = 3n,\quad  l_3 = 3n + 1,
\ee
or we can write
\be
l = 3n + (j-2)
\ee
where $j = 1, 2, 3$ and $n$ runs from 1 to $N$ with $N$ being the number 
of PGs in a hydrogen bond strand. 
 
Thus, for the ennumeration of PGs in an $\alpha$-helix, we can use 
the two numbers $j$ and $n$ where $j$, a cyclic index  modulo 3, 
indicates the spine of the hydrogen bond, and $n$ ennumerates PG in a 
spine or elementary cells of three PGs from different spines. We can 
use a different numbering of the cyclic index:  $j'= j-2 = -1, 0, 
+1$, or $j''= j-1 = 0, 1, 2$, or  $j = 1, 2, 3$. 

Introducing a double index $\{j,n\}$, instead of the single number $l$, 
the equilibrium positions of PGs in an $\alpha$-helix (\ref{aleqvpos}) can be 
rewritten as 
\begin{equation}
\vec{R}_{j,n}^{(0)} = r \left( \vec{e}_x \cos(\frac{2\pi n}{6}-\theta_j) - 
\vec{e}_y \sin(\frac{2\pi n}{6}-\theta_j)\right) + \vec{e}_z (\frac{5an}{6} +
\Delta_j),
\label{eqvpos}
\end{equation}
where $\theta_j=\frac{2\pi}{3.6}j-\theta_0$ and $\Delta_j=\frac{aj}{3.6}-z_0$. 
The spines of hydrogen bonds in  an $\alpha$-helix are also rolled into a 
helix of length $5a$ with $6$ PGs per turn. 

Due to the softness of hydrogen bonds, PGs can be displaced and their 
positions in an $\alpha$-helix are
\be
\vec{R}_{j,n} = \vec{R}_{j,n}^{(0)} + \vec{u}_{j,n} 
\ee
where $\vec{u}_n$ are the displacements of the peptide groups from their
equilibrium positions (\ref{eqvpos}).

The potential energy of displacements depends on the distance between 
the groups and so we can perform the approximation of using only the
nearest neighbours interaction. The nearest
neighbours along the polypeptide chain are bound together by 
rigid valence bonds, much more rigid than the hydrogen bond. We can 
thus assume that the distances between $l$-th and $(l \pm 1)$-th 
groups are fixed while the potential energy of displacements is 
determinded only by the variation of the hydrogen bond length and, 
in an harmonic approximation, it can be written as 
\begin{equation} 
{\cal{V}}=\sum_{j,n}[V(R_{j,n;j,n-1})-V(R_0)]=
\sum_{j,n}\frac{1}{2}w_H(\Delta R_{j,n;j,n-1})^2,
\label{potenen}
\end{equation}
where $w_H$ is the elasticity of the hydrogen bond. In (\ref{potenen}),
\begin{equation}
R_0=R_{j,n;j,n-1}^{0}=|\vec{R}_{j,n}^{(0)} - \vec{R}_{j,n-1}^{(0)}|=
\sqrt{(2r\sin(\pi/6))^2+(5a/6)^2},
\label{equilR}
\end{equation}
is the equilibrium length of the hydrogen bond, and 
\begin{equation}
\Delta R_{j,n;j,n-1} = |\vec{R}_{j,n} - \vec{R}_{j,n-1}| - R_0 =
\frac{(\vec{R}_{j,n} - \vec{R}_{j,n-1})(\vec{u}_{j,n} -\vec{u}_{j,n-1})}{R_0} 
\label{displace}
\end{equation}
are its changes due to the small displacements. 
The total energy of the displacements is the sum of 
the potential energy (\ref{potenen}) and the kinetic energy which is 
given by the relation
\begin{equation}
T=\sum_{j,n}\frac{1}{2}M \dot{\vec{u}}_{j,n}^2,
\end{equation}
where $M$ is the mass of a PG and 
$\dot{\vec{u}}_{j,n}=\frac{d\vec{u}_{jn}}{dt}$ 
are the velocities of the displacements.

Due to the assumption that the valence bonds are 
sufficiently rigid and that the 
distances 
between the $l$-th and $(l \pm 1)$-th groups are fixed, the three components 
of 
the PG displacement are not independent. In fact, we have two
 conditions  
which correspond to the assumption that the distances between each $l$-th PG 
and 
its two neighbours, $l-1$ and $l+1$, are fixed. For small displacements 
this means that the displacement $u_l$ of the $l$-th PG is orthogonal to the 
vectors connecting the  $l$-th and the $(l \pm 1)$-th groups:
\begin{equation}
\vec{u}_l\,\cdot \,(\vec{R}_l - \vec{R}_{l-1}) = 0,\qquad 
\vec{u}_l\,\cdot \,(\vec{R}_l - \vec{R}_{l+1}) = 0 .
\label{ortog}
\end{equation}
Let us represent the vector $\vec{u}_l$ using three orthogonal unit vectors 
$\vec{e}_{l}^{(r)}$, $\vec{e}_{l}^{(t)}$ and $\vec{e}_{z}$:
\be
\vec{u}_{l} = \vec{e}_{l}^{(r)} u_{l}^{(r)} + \vec{e}_{l}^{(t)}u_{l}^{(t)}
+ \vec{e}_{z}u_{l}^{\parallel},
\ee
where $\vec{e}_{z}u_{l}^{\parallel} = \vec{u}_{l}^{\parallel}$ is the 
longitudinal, (along the $\alpha$-helix axis) component of the 
displacement. The transversal component $\vec{u}_{l}^{\perp} = 
\vec{e}_{l}^{(r)} u_{l}^{(r)} + \vec{e}_{l}^{(t)}u_{l}^{(t)}$  is 
represented through the radial and tangential components relative to the axis. 
Here 
\begin{equation}
\vec{e}_{l}^{(r)}=\vec{e}_x \cos(\frac{2\pi l}{3.6}) + 
\vec{e}_y \sin(\frac{2\pi l}{3.6}),\qquad 
\vec{e}_{l}^{(t)}=-\vec{e}_x \sin(\frac{2\pi l}{3.6}) + 
\vec{e}_y \cos(\frac{2\pi l}{3.6}).
\label{rtorts}
\end{equation}
In this case condition (\ref{ortog}) takes the form
\begin{equation}
\frac{a}{3.6}u_{l}^{(||)} + 2r\sin^2(\frac{\pi}{3.6})u_{l}^{(r)} + 
r\sin(\frac{2\pi}{3.6})u_{l}^{(t)} = 0, \quad
\frac{a}{3.6}u_{l}^{(||)} - 2r\sin^2(\frac{\pi}{3.6})u_{l}^{(r)} + 
r\sin(\frac{2\pi}{3.6})u_{l}^{(t)} = 0.
\label{bondcond}
\end{equation}

From these equations it is easy to find that 
\begin{equation}
u_{l}^{(r)}=0, \qquad u_{l}^{(t)}=-\frac{a}{3.6 r\sin(2\pi/3.6)} 
u_{l}^{\parallel}.
\label{condis}
\end{equation}

Thus, there is only one independent degree of freedom of the PG displacement 
 and the vector of the displacement $\vec{u}_{j,n}$ can be represented 
as 
\be
\vec{u}_{j,n} = \vec{e}_{j,n} u_{j,n}
\ee
where 
\begin{equation}
\vec{e}_{j,n} = \frac{1}{C}\left[-a\left(\sin(\frac{2\pi n}{6}-\theta_j)
\vec{e}_x + \cos(\frac{2\pi n}{6}-\theta_j)\vec{e}_y \right) + 
3.6 r \sin(\frac{2\pi}{3.6}) \vec{e}_z \right],
\label{vecdisp}
\end{equation}
\be
C=\sqrt{a^2+\left(3.6\sin(\frac{2\pi}{3,6})r\right)^2}
\ee
is the unit vector which determines the direction of small displacements 
without changing of the valence bond length and $u_{j,n}$ is the amplitude of 
the displacements. 

Taking into account this expression, we obtain the following 
expression for the change of the hydrogen bond length 
(\ref{displace}) 
\begin{equation}
\Delta R_{j,n;j,n-1} = \gamma (u_{j,n} - u_{j,n-1}),
\label{hbchange}
\end{equation}
where 
\be
\gamma\,=\,\frac{ra}{CR_0}\left(\sin\frac{\pi}{3}\,+\,3\sin\frac{2\pi}{3.6}
\right).\label{gamma}
\ee

Thus, the potential energy (\ref{potenen}) is
\begin{equation}
{\cal{V}}=\sum_{j,n}\frac{1}{2}w(u_{j,n}-u_{j,n-1})^2,
\label{potenergy}
\end{equation}
where $w=\gamma^2 w_H$ is an effective elasticity coefficient. 

Therefore, the Hamiltonian of the $\alpha$-helix vibrations can be 
rewritten in the form
\begin{equation}
H_v = \sum_{j,n}\left[ \frac{p_{j,n}^2}{2 M} +
\frac{1}{2} w (u_{j,n} - u_{j,n-1})^2 \right]
\label{Hamvb3}
\end{equation}
where $p_{j,n}$ are the momentum operators that are canonically  
conjugate to the operators of the PG's displacement $u_{j,n}$.

We now focus on the Hamiltonian for the quasiparticle. 
The states of the Amid-I vibrations of
the peptide groups (or extra electron(s)) in the tight binding 
approximation are described by the Hamiltonian 
\begin{equation} 
H_e\ 
=\ \sum_{l}\left( E_0 A^+_{l}A_{l} + \sum_{m} L_{m}(A^+_{l}A_{l-m} + 
  A^+_{l-m}A_{l}) \right)
\label{Hamex1}
\end{equation}
where $l$ and $m$ run over the $3N$ values along the polypeptide chain. 
Here $\ A^+_{l}$ and $A_{l}\ $ are, respectively, the creation and annihilation operators of the 
quasiparticle at the $l$-th site of the chain; $L_m$ are 
the matrix elements of the excitation exchange between  
sites $l$ and $l \pm m$. The matrix elements $L_m$ with $m$ being a  
multiple of $3$ describe the energy exchange between the PGs of the 
same spine while
the others describe the excitation exchange between the spines. For Amid-I 
excitations in 
an $\alpha$-helix the numerical values of $L_m$ decrease with increasing $m$. 
In what follows we will take into account only two the most important terms: $L_1 = L$ which 
describes 
the interspine exchange, and $L_3 = -J$ which describes the 
intraspine one.  The signs of the corresponding matrix elements are 
chosen in such a way that they correspond to the polypeptide $\alpha 
$-helix \cite{Chirgadze,ASc}

Using the double index $\{ j,n \}$: $A_l = A_{j,n}$ 
we can rewrite (\ref{Hamex1}) as
\begin{eqnarray}
H_e & = & \sum_{n}\left[ \sum_{j}\left( E_0 A^+_{j,n}A_{j,n}\ - J
 A^+_{j,n}(A_{j,n+1} + A_{j,n-1}) \right) 
  \right. 
  +\nonumber\\ 
 & + & L [ A^+_{1,n}(A_{3,n-1} + A_{2,n})+ A^+_{2,n}(A_{1,n} + 
 A_{3,n}) +  A^+_{3,n}(A_{2,n} + A_{1,n+1})]  ]
\label{Hamex3}
\end{eqnarray}
where $n$ runs from 1 to $N$ and ennumerates the cells on each of the 
3 strands.

We now consider the Hamiltonian for the interaction of a quasiparticle with
the chain distortion. Due to the softness of the hydrogen bonds and 
the stiffness of the valence bonds, the distance between the $n$-th 
and $(n \pm 3)$-th group changes only under distortions of the 
$\alpha$-helix. So, taking into account the on-site deformation 
potential only (the dependence of $J$ on the distance between the groups 
is not so essential for an $\alpha$-helix \cite{ASc}), we can write 
the interaction Hamiltonian in the form 
\begin{equation} 
H_{int} = 
\sum_{j,n} \chi  ( u_{j,n+1} - u_{j,n-1}) A^+_{j,n}A_{j,n}.  
\label{Hint3}
\end{equation}
where $\chi$ is a constant 
parametrising the strength of the exciton(electron)-phonon interaction.

The total Hamiltonian
\begin{equation}
H = H_e + H_v + H_{int}
\label{Htot}
\end{equation}
where $ H_e$, $H_v$, and $H_{int}$ are given by (\ref{Hamex3}),
(\ref{Hamvb3}), and (\ref{Hint3}), respectively, describes the dynamics of a
molecular helical chain in which the equillibrium positions of its units 
(PGs) are given by the radius-vectors
\begin{equation}
\vec{R}_{j,n}^{(0)} = r \left( \vec{e}_x \cos(\frac{2\pi}{3}j) + \vec{e}_y
\sin(\frac{2\pi}{3}j)\right) + \vec{e}_z a (n + \frac{j}{3})
\label{eqvpos2}
\end{equation}
where $j$  is a cyclic index (of modulus $3$), which ennumerates the three
spines along the $z$ axes, and $n$ is the position index of an elementary cell 
within the
three strands. Equation (\ref{eqvpos2}) describes a molecular chain which is
rolled into a helix with three units per turn of the helix. We can 
consider such a molecular chain as a model of the $\alpha$-helical 
protein. In this case we can neglect the rolling of the hydrogen 
bonds into a superhelix.

\section{Elementary excitations in an $\alpha$-helix}

The quasiparticle Hamiltonian (\ref{Hamex3}) can be diagonalized by the 
following unitary
transformation
\begin{equation}
A_{j,m} = \frac{1}{\sqrt{N}} \sum_{\mu,k} e^{ikm} v_{j,\mu}(k) B_{\mu,k},
\qquad v_{j,\mu}(k) = \frac{1}{\sqrt{3}} e^{i(\mu + \frac{k}{3}) j},
\label{transf}
\end{equation}
where the wave number $k$ and the band index $\mu$ are given by
\begin{equation}
k = \frac{2 \pi}{N} l, \qquad l = 0, \pm 1, \ldots, \pm \frac{N-1}{2},
\qquad \mu = \frac{2 \pi}{3} \nu, \qquad \nu = 0, \pm 1.
\label{wnumb}
\end{equation}
Under transformation (\ref{transf}), the Hamiltonian (\ref{Hamex3})
transforms into
\begin{equation}
H_e = \sum_{\mu,k} E_{\mu}(k) B_{\mu,k}^+ B_{\mu,k}
\label{Hex}
\end{equation}
where the energy dispersion in the three bands ($\mu = 0, \pm \frac{2 \pi}{3}$)
is given by
\begin{equation}
E_{\mu}(k) = E_0 - 2J\cos (k) + 2L \cos \left(\frac{k}{3} +\mu \right)
\label{Eex}
\end{equation}
or, in explicit form, by
\begin{eqnarray}
E_0(k) & = & E_0 - 2J\cos (k) + 2L \cos \left(\frac{k}{3}\right) ,\nonumber\\
E_{\pm}(k) & = & E_0 -2J \cos (k) - L  \cos \left(\frac{k}{3}\right) \pm
\sqrt{3} L \sin \left(\frac{k}{3}\right) .
\label{3bands}
\end{eqnarray}

The Hamiltonian $H_v$ (\ref{Hamvb3}) describes independent oscillations of the 
PG in the spines of H-bonds in an $\alpha$-helix. 

Next, we perform a unitary transformation of the lattice variables: 
\be
 u_{jn} \,=\,\frac{1}{\sqrt{N}}\,\sum_q\,e\sp{iqn}\left(\frac{\hbar}
{2M\omega_q}\right)\sp{\frac{1}{2}}
(a_{j,q}\,+\,a_{j,-q}\sp{\dagger})
\ee
\be
p_{jn} \,=\,-\frac{i}{\sqrt{N}}\,\sum_q\,e\sp{iqn}\left(\frac{\hbar M\omega_q}
{2}\right)\sp{\frac{1}{2}}
(a_{j,q}\,-\,a_{j,-q}\sp{\dagger}),
\ee
where
$a_{j,-q}\sp{\dagger}$ and $a_{j,q}$ are the operators of creation and 
annihilation of acoustic phonons
with wavenumber $q$ and frequency  
\be
\omega_q=2\nu _a|\sin {\frac{q}{2}}|,\ \ \ \nu _a=\sqrt{\frac{w}{M}}.
\label{omega}
\ee

As it is convenient to describe the lattice oscillations in the helical 
 symmetry representation that we have introduced for the description 
of excitons, we define the operators $b_{\nu q}$ as
\be
a_{jq}\,=\,\sum_{\nu}\,v_{j\nu}(q)\,b_{\nu q},\ee
where the $v_{j\nu}(q)$ were given in the description of the excitons 
(\ref{transf}).
In this formulation, the displacement operator is given by
\be
u_{jn} \,=\,\frac{1}{\sqrt{N}}\,\sum_{q\nu}\,e\sp{iqn}
v_{j\nu}(q)\left(\frac{\hbar}{2M\omega_q}\right)\sp{\frac{1}{2}}
(b_{\nu,q}\,+\,b_{-\nu,-q}\sp{\dagger})\label{ubtransf}
\ee
and the Hamiltonian $H_v$ (\ref{Hamvb3}) takes the form
\be
H_v\,=\,\sum_{\nu q}\, \hbar \omega_q\,(b_{\nu,q}\sp{\dagger}b_{\nu,q}\,
+\,\frac{1}{2}).
\ee
Thus, the elementary excitations are given by the phonons which 
correspond to the deformational oscillations of the lattice, and the 
excitons which describe the internal Amid excitations of the PG. As 
an elementary cell contains 3 PGs, the spectrum consists of three 
exciton bands which correspond to the Davydov splitting. This band 
structure is shown in Fig. 1  for $L$=12.4 
cm$\sp{-1}$ and $J$=7.8cm$\sp{-1}$, which correspond to the 
$\alpha$-helix values.

%HERE WE PUT IN THE PICTURE OF THESE THREE FORMULAE (Bernard's plot of 31)

\begin{figure}[htbp]
\unitlength1cm \hfil
\begin{picture}(8,8)
 \epsfxsize=8cm \epsffile{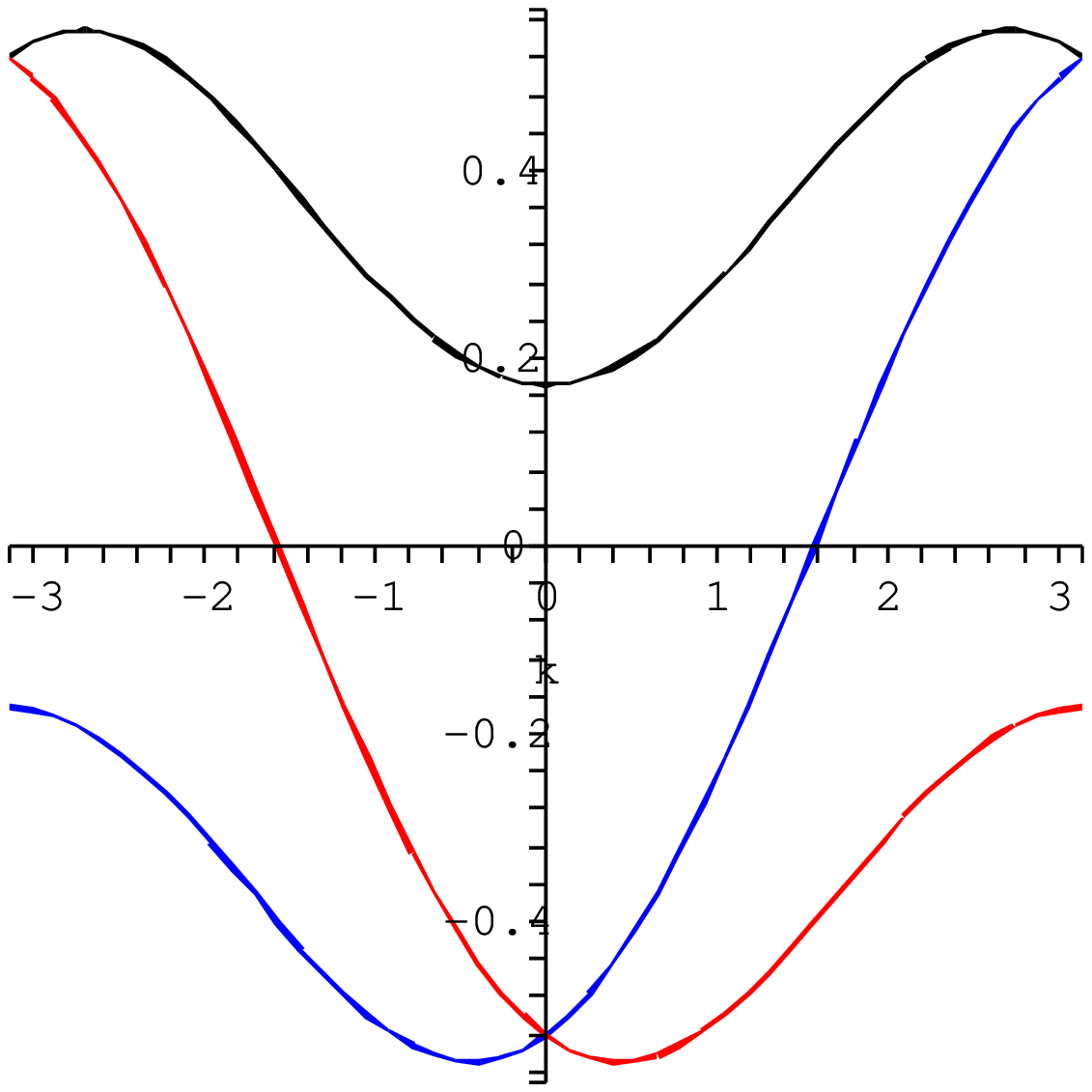}
 \put(-2.3,7){$E$}
 \put(-1.5,5){$E_+$}
 \put(-1.8,2){$E_-$}
\end{picture}
\caption{The three energy bands (\ref{3bands}) for $J=7.8cm^{-1}$ and 
$L=12.4cm$ }
\end{figure}

Finally, we rewrite the  interaction Hamiltonian $H_{int}$  
(\ref{Hint3}) as
\be
H_{int}\,=\,\frac{1}{\sqrt{3N}}\,\sum_{kq\mu\nu}\,\{\chi(q)B_{\mu+\nu,k+q}
\sp{\dagger}B_{\mu,k}b_{\nu,q}\,+\,
\chi\sp{\ast}(q)B_{\mu,k}\sp{\dagger}B_{\mu+\nu,k+q}b\sp{\dagger}_{\nu,q},\}
\ee
where
\be
\chi(q)\,=\,i\chi\left(\frac{2\hbar}{M\omega_q}\right)\frac{1}{2}\sin(q).
\label{chi}
\ee

%Section 4.

\section{Equations in the adiabatic approximation}

In the adiabatic approximation the wavefunction of the system with 
one quasiparticle  is represented as 
\be
|\psi(t)\rangle = U(t)|\psi_e(t)\rangle ,
\label{adappr}
\ee
where $U(t)$ is the unitary operator of the coherent molecule 
displacements
\be
U(t)\ =\exp{\left [\sum_{\nu,q 
}(\beta_{\nu,q}(t)b\sp{\dagger}_{\nu,q}\ -\ 
\beta\sp{\ast}_{\nu,q}(t)b_{\nu,q})\right ]}, 
\label{uoperat}
\ee 
\be
|\psi_e(t)\rangle=\sum_{\mu,k}\psi _{\mu,k}(t)B\sp{\dagger}_{\mu,k}|0\rangle, 
\label{grf}
\ee
with functions $\psi _{\mu,k}(t)$ that satisfy the normalisation 
condition:
\be
\sum_{\mu,k} |\psi _{\mu,k}(t)|^2=1.
\label{norm}
\ee
The coefficients $\beta_{\nu,q}(t)$ in  (\ref{uoperat}) are, at this 
stage, arbitrary functions which will be determined below.

In the adiabatic approximation the equations for  $\psi _{\mu,k}(t)$ 
and $\beta_{\nu,q}(t)$ can be obtained either directly from the time dependant
Schr{\"o}dinger equation or as 
Hamilton equations for the generalized variables $\psi _{\mu,k}(t)$, 
$\beta _{\nu,q}(t)$ and their canonically conjugated momenta 
$(-i/\hbar)\psi _{\mu,k}^{*}(t)$ and $(-i/\hbar)\beta 
_{\nu,q}^{*}(t)$ by considering 
\begin{eqnarray} 
{\cal{H}}&=&\langle \psi| H |\psi\rangle = 
\sum_{\mu,k} E_{\mu}(k) \psi_{\mu,k}^* \psi_{\mu,k} 
+ \sum_{\nu,q}\hbar \omega_q (\beta_{\nu,q}^*\beta _{\nu,q}
+ \frac{1}{2})\nonumber\\
&+& \frac{1}{\sqrt{3N}}\sum_{\mu,k,\nu,q} \chi (q) 
\psi_{\mu+\nu,k+q}^* \psi_{\mu,k} (\beta _{\nu,q} + \beta _{-\nu,-q}^*)
\label{Hamfunc}
\end{eqnarray}
as a Hamilton functional. The equations are thus given by 
\be
i\hbar\frac{d\psi _{\mu ,k}(t)}{dt}=E_{\mu }(k)\psi _{\mu ,k}(t)+
\sum_{q,\nu} \frac{2i\chi \sin q}{\sqrt{3NM}}  
Q_{\nu}(q,t)\psi _{\mu-\nu ,k-q}(t),
\label{eqpsi}
\ee
\be
i\hbar\frac{d\beta _{\nu,q}}{dt}=\hbar \omega_{q} \beta _{\nu ,q} +
\frac{1}{\sqrt{3N}} \sum_{\mu,q} \chi^* (q) \psi_{\mu,k}^*
\psi_{\mu+\nu,k+q}.
\label{eqbeta}
\ee
In the first equation, $Q_{\nu}(q,t)$ is given by
\be
Q_{\nu}(q,t)\,=\,\left(\frac{\hbar}{2\omega_q}\right)\sp{\frac{1}{2}}
\left(\beta_{\nu,q}\,+\,\beta\sp{\ast}_{-\nu,-q}\right).
\ee

In fact, the equation for $\beta_{\nu,q}(t)$ becomes the equation for
$Q_{\nu}(q,t)$ and takes the form:
\be
\frac{d^2Q_{\nu}(q,t)}{dt^2}+\omega^2_qQ_{\nu}(q,t)=\frac{2i\chi 
\sin{q}}{\sqrt{3NM}} \sum_{k,\mu }\psi ^*_{\mu ,k}(t)\psi _{\mu+\nu 
,k+q}(t).
\label{eqQ}
\ee
In these expressions, and in what follows, the index $\nu$ labeling 
$\psi$ and $Q$ is defined modulo 3.

Next, we seek the stationary solutions of these equations by
requiring that 
\be
\psi_{\nu,k}(t)\,=\,e\sp{-i(\Theta (t)+kz_s(t))}\,\psi_{\nu}(k).
\label{stpsi}
\ee
This immediately tells us that
\be
Q_{\nu}(k,t)\,=\,e\sp{-ikz_s(t)}\,Q_{\nu}(k).
\ee
Here the parameter $z_s(t)$ corresponds to the centre of mass 
of the excitatiton.

Substituting this ansatz into our equations, we obtain
\be
\left[\hbar \Omega + \hbar V k - E_{\mu}(k)\right]\psi _{\mu}(k) =
\sum_{\nu,q} \frac{2i\chi \sin q}{\sqrt{3NM}} Q_{\nu}(q)\psi
_{\mu-\nu}(k-q),
\label{steqpsi}
\ee
\be
(\omega^2_{q} - V^2q^2)Q_{\nu}(q,t)=\frac{2i\chi
\sin{k}}{\sqrt{3NM}} \sum_{k,\mu }\psi ^*_{\mu}(k)\psi 
_{\mu+\nu}(k+q),
\label{steqQ}
\ee
where $\Omega = \frac{d\Theta}{dt}$ and $V=\frac{dz_s}{dt}$ is the velocity of
the propagation of the excitation measured in units of the lattice 
constants.

Taking into account (\ref{steqQ}), we see that (\ref{steqpsi}) is a
nonlinear integral equations. From (\ref{steqpsi}) we see that 
$\psi_{\mu}(k)$ has a maximum at the carrying wave number $k_c$ which
corresponds to the minimum of $\hbar \Omega + \hbar V k - E_{\mu}(k)$, i.e.
$k_{c\mu}$ and the excitation velocity $V$ are connected by the relation
\be
\hbar V = \frac{d E_{\mu}(k)}{d k} \vert_{k=k_{c\mu}} .
\label{carvel}
\ee

Next, we assume that, in the space representation, the solution is 
given by a wave packet broad enough so that it is sufficiently narrow 
in the $k$ representation. This means that $\psi_{\mu}(k)$ are 
essentially nonzero only in a small region of values of $k$ in the 
vicinity of $k_{c\mu}$. In this case we can use the following 
approximation
\be
\hbar \Omega + \hbar V k - E_{\mu}(k) = \Lambda -\frac{\hbar
^2 (k-k_{c\mu})^2}{2m_{\mu}},
\label{enappr0}
\ee
where
\be
\Lambda = \left[ \hbar \Omega + \hbar V k - 
E_{\mu}(k)\right]_{k=k_{c\mu}}, \qquad 
\frac{\hbar^2}{m_{\mu}} = 
\frac{d^2E_{\mu}(k)}{dk^2}\vert_{k=k_{c\mu}}.
\label{efmass}
\ee

To solve  (\ref{steqpsi}), we introduce
the position dependent functions 
\be
\varphi_{\mu}(x)=\frac{1}{\sqrt {N}} \sum_{k} e^{i(k-k_{c\mu})x} 
\psi_{\mu}(k).
\label{varphi}
\ee
Note that at $x=n$ this is a unitary transformation of $\psi_{\mu}(k)$
to the site representation. Using approximation (\ref{enappr0}), one 
can transform  (\ref{steqpsi}) into a differential equation for 
$\varphi_{\mu}(x)$: 
\be
\Lambda \varphi _{\mu}(x) + \frac{\hbar^2}{2m_{\mu}} \frac{d^2\varphi 
_{\mu}(x)}{d x^2} - \sum_{\nu}{\cal{V}}_{\nu}(x) 
e^{-i(k_{c\mu}-k_{c(\mu -\nu)})x} \varphi _{\mu-\nu}(x) = 0, 
\label{eqvarphi}
\ee
where 
\be
{\cal{V}}_{\nu}(x) = \frac{i\chi}{\sqrt{3MN}} \sum_{q} e^{iqx}
Q_{\nu}(q) \sin{q}.
\label{varV}
\ee
Note also that (\ref{eqvarphi}) is only a zero-order approximation of 
(\ref{steqpsi}) and so it corresponds to the continuum approximation. 
Only in this approximation the soliton velocity $V$ and frequency  
$\Omega$ are constant and the soliton centre of mass evolves with time as 
$z_s(t) = Vt + z_0$ and $\Theta (t) = \Omega t$. 

Finally note that, when transforming (\ref{steqpsi}) into (\ref{eqvarphi}), one
has to be careful with the  double summation
($\sum_{k,q}...$).  The wavenumbers $k$ and $q$ are in the first
Brillouin zone, $-\pi <\ k\ ,q\ \leq\ \pi$, and the wave number $k-q$
 also has to be in this zone.  This is the case for small
values of $k$ and $q$ (normal processes in exciton-phonon
interactions). However, when $k$ and $q$ are close to the edge of the first
Brillouin zone, it is possible that $|k-q| > \pi$ (Umklapp processes).
In this case it is nessesary to reduce the wavenumber $k-q$ to the first
Brillouin zone using the reciprocal lattice wavenumber $g=2\pi$.
This does not change the discrete equations due to the periodicity
of the functions in the space of reciprocal lattice vectors, but is
essential when introducing continuous functions for the analytical
investigations. The Umklapp processes lead to the appearence of additional
terms in (\ref{eqvarphi}) for which the double summations are
performed in the regions near the edges of the Brillouin zone where 
$|q-k|>\pi$.  The assumption that $\psi_{k}$ and $Q(q)$ are small in 
these regions allows  us to consider these terms as a perturbation.  Here 
we do not take this perturbation into consideration. A detailed analysis 
can be found in \cite{dyn} where it has been shown that allowing Umklapp 
processes in $k$ space leads to the appearence of a periodical (with a 
period of a lattice constant) Peierls-Nabarro potential barrier for the
motion of the soliton centre of mass (\cite{Kupr,Vakh,Vakh1,dyn}). As a 
result, in discrete lattices, the ``instantaneous'' soliton velocity 
depends on time and has an oscillatory component with a period 
\be 
T_d = \frac{2\pi}{V_{av}}, 
\label{pierlsperiod}
\ee
where $V_{av}$ is the average velocity of the soliton propagation in 
the chain.

Having found the solutions of (\ref{eqvarphi}), we  can then use 
the transformation (\ref{transf}) and, taking into account (\ref{stpsi}) 
and (\ref{varphi}), write down the  probability amplitudes for 
the distribution of the excitations in an $\alpha$-helix:
\begin{equation}
\Psi_{j,n}(t) = \frac{1}{\sqrt{N}} \sum_{\mu,k} e^{ikn} v_{j,\mu}(k) 
\psi_{\mu,k}(t) = \frac{1}{\sqrt{3}}\sum_{\mu} e^{-i(\Omega + 
Vk_{c\mu})t +ik_{c\mu}n + i(\mu + \frac{1}{3}k_{c\mu})j} 
\varphi_{\mu}(n+\frac{1}{3}j - Vt - z_0).
\label{siterep} 
\end{equation} 

From (\ref{steqQ}) we obtain the explicit expressions for 
$Q_{\nu}(q)$ at $\nu = 0$ and $\pm 2\pi/3$, namely:
\be
Q_{0}(q)=\frac{2i\chi \sin{q}}{(\omega _q^2-V^2q^2)\sqrt{3MN}} 
\sum_{\mu ,k} \psi ^*_{\mu ,k}\psi _{\mu ,k+q},
\label{Q0}
\ee
\be
Q_{+}(q)=\frac{2i\chi \sin{q}}{(\omega _q^2-V^2q^2)\sqrt{3MN}} 
\sum_{k}\left( \psi ^*_{0}(k)\psi _{+}(k+q) +
 \psi ^*_{+}(k)\psi _{-}(k+q) + \psi ^*_{-}(k)\psi _{0}(k+q) \right),
\label{Qnu}
\ee
\be
Q_{-}(q) = Q_{+}^{*}(-q).
\ee

Substituting these expressions into (\ref{varV}), we obtain the  
potentials ${\cal{V}}_{\nu}(x)$. For example,  
\be 
{\cal{V}}_{0}(x) = - 
\frac{1}{N} \sum_{q,k,\mu} \frac{2\chi^2 \sin^2{q}}{3M(\omega 
_q^2-V^2q^2)} e^{-ikx}\psi ^*_{\mu}(k) e^{i(k+q)x} \psi _{\mu}(k+q) .  
\ee

We can see from (\ref{Q0}) that $Q_0(q)$ is essentially nonzero only at 
small values of $q$. So we can use the long-wave approximation 
and write the phonon dispersion relation (\ref{omega}) as  
$\omega_{q} \approx \nu _a |q| $. 
In this case ${\cal{V}}_0(x)$  is given by
\be
{\cal{V}}_{0}(x) = - \frac{2\chi^2 }{3w(1-v^2)}\sum_{\mu} 
|\varphi_{\mu}(x)|^2 
\label{calV0}
\ee
where $v = V/\nu_a$ is a velocity in units of sound velocity $\nu_a$.
Similarly, the potentials ${\cal{V}}_{\pm}(x)$ are quadratic in
$\varphi_{\mu}(x)$. Therefore, the system of  equations (\ref{eqvarphi}) is 
a system of nonlinear Schr\"{o}dinger equations (NLSEs).

We observe that equations (\ref{eqvarphi})
admit three types of ground state solutions of a soliton type which 
preserve the helical symmetry of the system. Such solutions describe 
solitons which are formed by excitons from only one of the three 
excitonic bands, i.e. only one function $\varphi _{\mu}\neq 0$ for a
given $\mu$ is nonzero and the other two  $\varphi _{\nu }=0$ with $\nu 
\neq \mu$. In such states, according to (\ref{Q0})-(\ref{Qnu}), only 
the total symmetrical distortion of the $\alpha$-helix takes place, i.e. 
$Q_{0}(q) \neq 0$ and $Q_{\pm}(q)=0$. Taking into account 
(\ref{calV0}), we note that these types of solitons are described by 
the NLSE: 
\be
\Lambda \varphi _{\mu}(x) + \frac{\hbar^2}{2m_{\mu}} \frac{d^2\varphi 
_{\mu}(x)}{d x^2} + \frac{2\chi ^2}{3w(1-v^2)}|\varphi _{\mu}(x)|^2 
\varphi _{\mu}(x) = 0 
\label{nls1} 
\ee
together with the normalisation condition (\ref{norm}). Its solution is given 
by
\be
\varphi _{\mu}(x)=\sqrt{\frac{\kappa_{\mu} }{2}} \frac{1}{\cosh 
(\kappa_{\mu} x)}  
\label{fmu}
\ee
with the eigenvalue
\be
\Lambda_{\mu} = - \frac{\hbar^2 \kappa _{\mu}^2}{2m_{\mu}},
\ee
where
\be
\kappa _{\mu}=\frac{m_{\mu}\chi ^2}{3\hbar w(1-v^2)}.
\label{kappa}
\ee
Thus, from (\ref{efmass}) we find that
\be
\hbar \Omega = E_{\mu}(k_{c\mu}) - Vk_{c\mu} - \frac{\hbar^2 \kappa 
_{\mu} ^2}{2m_{\mu}} .
\label{Omegmu} 
\ee
According to (\ref{siterep}),
\begin{equation}
\Psi_{j,n}(t) = \sqrt{\frac{\kappa_{\mu} }{6}} \frac{e^{-i(\Omega_{\mu} + 
Vk_{c\mu})t +ik_{c\mu}(n+ \frac{1}{3}j) + i\mu j}}
{\cosh {\kappa_{\mu} (n+\frac{1}{3}j - Vt - z_0)}}.
\label{siterepmu} 
\end{equation} 
This excitation is spatially distributed between the chains with the 
probability components given by:
\be
P _{j,n}(t)=\frac{1}{3} \varphi _{\mu}^2(n+\frac{j}{3}-Vt-z_0).
\label{prmu}
\ee
Clearly, $P_j = \sum_{n}P_{j,n} = 1/3$. For the totally symmetric 
soliton, $\mu = 0$, the chains are excited with the same phase, while 
for the other two cases, $\mu = \pm 2\pi /3$ and  the excitations in the  
spines have the phase shifts $\pm 2\pi /3$.

Note that due to the factor $(1-v^2)$ in (\ref{calV0}) and
(\ref{kappa}) we see that the soliton velocity $V$
cannot exceed the sound velocity $\nu_a$. However,  there is also
a further  restriction on the soliton velocity which follows from
 (\ref{carvel}). Unlike for the parabolic law, the energy dispersion
in an exciton band shows that $dE(k)/dk$ has a maximum value.
Therefore, (\ref{carvel}) has a solution only when $V$ does not
exceed the maximum exciton group velocity $V_g =
(1/\hbar)(dE(k)/dk)_{max}$ and so, the top speed of the solitons is
determined by the lowest of $\nu_a$ and $V_g$.  For example, in a 
simple chain with $E(k) = - 2J \cos{k}$, $V_g = 2J/\hbar$, and with  
the parameters of the $\alpha$-helix, $\nu_a > V_g$.

Below we will consider solitons at low velocities. In this case, from 
(\ref{carvel}), we have
\be
k_{c\mu} = k_{\mu} + \frac{m_{\mu}}{\hbar}V .
\label{kcmu}
\ee
 Here  $k_{\mu}$ determines the 
bottom of the $\mu$-th exciton band and $m_{\mu}$ is an effective exciton 
mass near the band bottom. At low velocities the total energy 
${\cal{E}}_{\mu} = {\cal{H}}$ of the soliton state is given by
\be
{\cal{E}}_{\mu}(V) = {\cal{E}}_{\mu}(0) + \frac{1}{2}{\cal{M}}_{\mu} 
V^2.
\label{Esolmu}
\ee

The totally symmetric exciton band has a minimum at $k_0 = 0$ and, in 
the long-wave approximation, we have 
\be
E_0(0) = E_0-2J+2L, \qquad m_0=\frac{9\hbar^2}{2(9J-L)} 
\label{E-m0}
\ee
Therefore, the totally symmetric soliton state is characterized by 
the width parameter 
\be 
\kappa_0 =\frac{3\chi ^2}{w(9J-L)}, 
\label{kappa0} 
\ee
the energy
\be
{\cal{E}}_0(0) = E_0-2J+2L-\frac{\chi ^4}{3w^2(9J-L)},
\label{erest0}
\ee
and the mass
\be
{\cal{M}}_{0}= m_0+\frac{8\chi ^4}{3\nu _a^2 w^2 (9J-L)}.
\label{msol0}
\ee

The other two of the three soliton states are formed by excitons 
from the other two  bands, $\mu = \pm 2\pi/3$. Due to the helical 
symmetry, the bottoms of these bands are determined by the non-zero  
wavenumbers $k_{\pm} = \pm k_d$. For an $\alpha$-helix the parameter 
$k_d$ is small and can be determined in the long-wave approximation 
as 
\be 
k_d=\frac{9L}{\sqrt{3}(18J+L)}.  
\label{k*} 
\ee 
Thus, for these 
bands we have 
\be 
E_{\pm}(0)=E_1 = E_0-2J-L-\frac{3L^2}{2(18J+L)},  
\qquad m_{\pm}=\frac{9\hbar^2}{18J+L} \equiv m_1, 
\label{E-mpm} 
\ee

Therefore, these soliton states are characterized by the
width parameter 
\be
\kappa _1=\frac{6\chi ^2}{w(18J+L)(1-v^2)}, 
%s^2=\frac{v^2}{\nu _a^2},
\label{kappa1}
\ee
the total energy at rest
\be
{\cal{E}}_1(0) = E_0-2J-L-\frac{3L^2}{2(18J+L)} -\frac{2\chi 
^4}{3w^2(18J+L)}, 
\label{erest1} 
\ee 
and by the soliton mass 
\be 
{\cal{M}}_{1}= 
m_1+\frac{16 \chi ^4}{3(18J+L)w^2 \nu _a^2}.
\label{msol1} 
\ee

Note that the energies of these three solitons are split from the 
bottoms of the corresponding energy bands. We should add that the 
solutions of these soliton states were also found in \cite{FedYak}.

The energy levels of the last two solitons are degenerate. However, 
according to the Jan-Teller theorem, this degeneracy can be broken by 
the distortions of the chains and a hybridization of these two 
states can take place. Below we consider such a case when 
i.e., $\varphi _{\pm} \neq 0 $ and $\varphi _0 =0$. 

In this case we find from (\ref{eqvarphi}) that $\varphi _{\pm}$ are 
determined by the system of equations:
\be
\Lambda \varphi _{+}(x) + \frac{\hbar^2}{2m_{1}} \frac{d^2\varphi 
_{+}(x)}{d x^2} - {\cal{V}}_{0}(x)\varphi _{+} - 
e^{-i(k_{+}-k_{-})x}{\cal{V}}_{-}(x) \varphi _{-}(x) = 0,
\label{eqvarphi1}
\ee
\be
\Lambda \varphi _{-}(x) + \frac{\hbar^2}{2m_{1}} \frac{d^2\varphi 
_{-}(x)}{d x^2} - {\cal{V}}_{0}(x)\varphi _{-} - 
e^{i(k_{+}-k_{-})x}{\cal{V}}_{+}(x) \varphi _{+}(x) = 0.
\label{eqvarphi2}
\ee
In this case, the components $Q_{\pm}$ of the deformation of the 
$\alpha$-helix are also non-zero:
\be
Q_{+}(q)=\frac{2i\chi \sin{q}}{(\omega _q^2-V^2q^2)\sqrt{3MN}} 
\sum_{k} \psi ^*_{+}(k)\psi _{-}(k+q).
\label{Qp}
\ee
Substituting this into (\ref{varV}) we find that, for small velocities,
\be
{\cal{V}}_{+} = - \frac{2\chi^2 }{3w)} e^{-i(k_{+}-k_{-})x} 
\varphi_{+}^*(x) \varphi _{-}(x), \quad 
{\cal{V}}_{-} = {\cal{V}}_{+}^{*}.
\label{calVp}
\ee
The deformational potential ${\cal{V}}_0$ of the totally symetric 
distortions is given by
\be
{\cal{V}}_{0}(x) = - \frac{2\chi^2 }{3w}\left( |\varphi_{+}(x)|^2 + 
|\varphi_{-}(x)|^2 \right).
\label{calV01}
\ee
Thus, equations (\ref{eqvarphi1}) and (\ref{eqvarphi2}) give us a 
system of NLSEs:
\be
\Lambda \varphi _{+}(x) + \frac{\hbar^2}{2m_{1}} \frac{d^2\varphi 
_{+}(x)}{d x^2} + \frac{2\chi^2}{3w}\left( |\varphi _{+}|^2 + 
2|\varphi _{-}|^2 \right) \varphi _{+}(x) = 0 
\label{sysnlse} 
\ee
and, equivalently,  for $\varphi _{-}$.

The general solution of these equations, normalized by the condition (\ref{norm}), 
is 
\be
\varphi _{\pm} = \frac{1}{\sqrt{2}} e^{\theta_{\pm}}\varphi _{2},
\ee
where $\theta_{\pm}$ are arbitrary phases and $\varphi _{2}$ 
satisfies the NLSE and is, therefore, given by (\ref{fmu}) with
\be
\kappa _2=\frac{9\chi ^2}{w(18J+L)}.
\label{kappa2}
\ee
The total energy of this soliton state at $V=0$ is
\be
{\cal{E}}_2(0) = E_0-2J-L-\frac{3L^2}{2(18J+L)} -\frac{3\chi 
^4}{2w^2(18J+L)}, 
\label{erest2} 
\ee  
and the soliton mass is
\be 
{\cal{M}}_{2}= 
m_1+\frac{12 \chi ^4}{w^2 \nu _a^2(18J+L)}.  
\label{msol2} 
\ee

Representing the energies of the other two solitons  
(\ref{erest1}) in the form ${\cal{E}}_1(0)=E_{b}-\Delta $ with 
$E_b = E_{\pm}(k_d)$ being the corresponding bottom of the energy 
band and 
\be
\Delta = \frac{2 \chi ^4}{3w^2(18J+L)},
\label{del}
\ee
we can write
\be
{\cal{E}}_2(0)=E_b(0)-\frac{9}{4}\Delta =
{\cal{E}}_1(0)-\frac{5}{4}\Delta .
\label{en2}
\ee
Thus, we see that the latter hybrid soliton has the lowest 
energy.

The distribution of the excitation amongst the chains 
is given by the probability amplitude:  
\be 
\psi^{(h)}_{j,n}(t)=\sqrt{\frac{2}{3}} e\sp{-i(\Omega t - 
\frac{\hbar}{m_1}V(n+j/3) - \frac{\theta _{+} + \theta _{-}}{2})}
\cos \left(k_d(n+j/3) + \frac{\theta _{+} - \theta _{-}}{2} + 
\frac{2\pi }{3}j\right) \varphi _2(n+\frac{j}{3},t).  
\label{distr3} 
\ee
Therefore,
\be 
P_{j,n} = |\psi^{(h)}_{j,n}(t)|^2 = \frac{\kappa_2}{3}
\frac{1 + \cos \left(2k_d(n+j/3) + (\theta _{+} - \theta _{-}) - 
\frac{2\pi }{3}j\right)}{\cosh ^2(n+\frac{j}{3}-Vt-z_0)}.  
\label{dpr3} 
\ee

Next, we consider the probability distribution of the excitation 
summed over all the spines of the helix:
\be
P_n(t)=\sum_{j}P_{j,n}(t) = |\varphi _2 (n,t)|^2 \left(1 - 
\frac{k_d}{3 \sqrt{3}} \cos (2k_dn + \theta _{+} - \theta _{-}) 
\right) \approx |\varphi _2 (n,t)|^2
\label{probs}
\ee
for small $k_{d}$, and the total probability of the excitation 
localisation on a given spine:
\be
P_j(t)=\sum_{n}P_{j,n}(t)|^2 = 
\frac{1}{3}\left[1-\frac{\pi k_d}{\kappa _2 \sinh \frac{\pi 
k_d}{\kappa _2 }} \cos \left(2k_dVt - \frac{2\pi }{3}j + \theta 
_{+} - \theta _{-} \right)\right].  
\label{probch} 
\ee 
We see from   (\ref{probch}) that 
the probability of the excitation localisation on a given spine is 
an oscillatory function of time with the period of oscillations given 
by 
\be 
T=\frac{\pi }{k_d V}.  
\label{period} 
\ee

Thus, the helical symmetry of the system results in the 
interspine soliton oscillations with a period of oscillations that  
is determined by the soliton velocity and the quasimomentum value 
corresponding to the bottom of the energy band (\ref{k*}).
These oscillations get mixed up with the oscillations that arise from the 
influence of the lattice discretness on the soliton dynamics which leads to 
the appearance of the Peierls-Nabarro potential. The period of these 
latter oscillations is also determined by the soliton velocity, 
(\ref{pierlsperiod}), as is shown in \cite{dyn}.

\section{Numerical modeling}

For the numerical calculations we consider an $\alpha $-helical system of 
length $N=150$ with periodic boundary conditions:
\be 
f_{j,n+N}=f_{j,n}
\ee
or, equivalently, 
\be
f_{l+3N}=f_{l},
\ee
where $f$ stands for $\psi$ or $\beta $ and the index $l$ ennumerates 
the sites along the polypeptide chain and $\{j,n\}$ denotes the site number 
$n$ in the $j$-th hydrogen bound spine ($j=1,2,3$).

It is more convenient, for the numerical simulations,  to use the
physically more relevant site representation for the $\Psi_{j,n}$ 
variables and to use $u_{j,n}$ for the displacements of PGs from the 
positions of their equilibrium. Here $u_{j,n}$ are the average 
displacements of PGs in the state (\ref{adappr}) and are related to 
$\beta_{\nu,q}$ by the unitary transformation (\ref{ubtransf}).  In 
these variables the equations (\ref{eqpsi})-(\ref{eqbeta}) 
become 

$$i\hbar \frac{d\Psi _{1,n}}{dt}=E_0\Psi _{1,n}-J(\Psi 
_{1,n-1}+\Psi _{1,n+1}) + L(\Psi_{3,n-1} +\Psi _{2,n}) + $$
\be 
\chi (u_{1,n+1}-u_{1,n-1}) \Psi _{1,n}, 
\label{-1} 
\ee 
$$i\hbar \frac{d\Psi _{2,n}}{dt}=E_0\Psi _{2,n}-J(\Psi _{2,n-1}
+\Psi _{2,n+1}) + L(\Psi_{1,n} +\Psi _{3,n}) + $$
\be 
\chi (u_{2,n+1}-u_{2,n-1}) \Psi _{2,n}, 
\label{0} 
\ee 
$$i\hbar \frac{d\Psi _{3,n}}{dt}=E_0\Psi 
_{3,n}-J(\Psi _{3,n-1}+\Psi _{3,n+1}) + L(\Psi_{2,n} +\Psi 
_{1,n+1}) + $$
\be 
\chi (u_{3,n+1}-u_{3,n-1}) \Psi _{3,n}, 
\label{+1} 
\ee 
\be
M\frac{d^2 u _{j,n}}{dt^2}=-w(2u _{j,n}-u _{j,n-1} -u 
_{j,n+1})+\chi (|\Psi _{j,n+1}|^2-|\Psi _{j,n-1}|^2),\ \ \ j=1,2,3.
\label{beta}
\ee
For $n=0$ and $n=N-1$ in the expressions above we take the appropriate values 
of the functions determined by our periodicity conditions.

In our studies we have adopted the following procedure.  We have 
started off with a reasonable field configuration and then used it as 
an initial excitation to determine a stationary solution of our 
system of equations (\ref{-1}-\ref{beta}).  Having determined this 
solution  numerically, we have kept on modifying it by an adiabatical 
increase of the wave-vector (thus increasing the velocity of the 
soliton), and have found for each fixed value of the wave-vector the 
corresponding stationary solution describing a soliton which 
propagates along the helix with an increasing non-zero velocity, 
determined by the gradually increasing values of the carrying 
wave-vector.

Note that, for an $\alpha$-helix, the question about the initial configuration is 
more important than for a simple chain, because there are three types 
of solutions, corresponding to the different symmetries. Studying similar 
problem for a simple linear chain, we had two equivalent approaches 
of deriving a stationary solution at zero velocity (see, e.g., 
\cite{dyn}).  Namely, we could start with the system of stationary 
equations and find the solution by minimizing the energy using 
some standard procedures. Another approach would use 
the non-stationary equations  which include some dissipation of the 
energy  in the lattice sub-system.  Starting with an arbitrary 
localized initial configuration of an excitation, we would find some 
time later
a stationary solution at zero velocity. Then
this configuration would be modified by adding a small carrying wave-vector 
and would be used as a starting initial condition for the next set of
calculations of the system of equations without any dissipation. 
This would result in a solution moving with a small non-zero velocity.  
Repeating this procedure further, we would increase the soliton velocity 
adiabatically till the velocity reaches the maximum value 
corresponding to the chosen parameters of the chain. 
This last approach, of using an arbitrary initial configuration, in the case 
of a helical structure probably cannot describe all possible
 solutions, since it would always lead to the solution of the 
lowest energy.

The energy expression can be obtained from   (\ref{Hamex3}, \ref{Hamvb3}, 
\ref{Hint3}). In the site representation the total energy is given by
\be
E_{tot}=E_e+E_v+E_{int},
\label{entot}
\ee 
where
\begin{eqnarray}
E_e & = & \sum_{n=0}^{N-1}\left[ \sum_{j=1}^{3}\left( E_0 \cmod 
\Psi_{j,n}\ - J (\Psi^*_{j,n}\Psi_{j,n-1} + \Psi^*_{j,n-1}\Psi_{j,n}) 
\right) \right.  + \nonumber\\ 
& + & L ( \Psi^*_{1,n}\Psi_{3,n-1} + 
\Psi ^*_{3,n-1}\Psi_{1,n} + \Psi^*_{2,n}\Psi_{1,n} +  
\Psi^*_{1,n}\Psi_{2,n} +  
  \nonumber\\ 
 & + & \left. \Psi^*_{3,n}\Psi_{2,n} + \Psi^*_{2,n}\Psi_{3,n}) 
 \right], 
 \label{ene} 
 \end{eqnarray}
\begin{equation}
E_v = \sum_{j=1}^{3} \sum_{n=0}^{N-1}\left[ \frac{M}{2} 
\left(\frac{d u _{j,n}}{dt}\right)^2 + \frac{1}{2} w (u_{j,n} 
- u _{j,n-1})^2 \right],
\label{env}
\end{equation}
\begin{equation}
E_{int} = \sum_{j=1}^{3} \sum_{n=0}^{N-1} \chi  ( u _{j,n+1} - 
u _{j,n-1}) \cmod \Psi_{j,n}.   
\label{enint} 
\end{equation}

In our simulations we have taken the
 numerical values of the parameters from \cite{Sct1}: i.e.
$L=12.4 cm^{-1},\ J=7.8 
cm^{-1}$, $\chi = .34 \cdot 10^{-10}$ N, $w=19.5$ N/m and 
$\sqrt{M/w}=1/\nu_a=0.99 \cdot 10^{-13}$ s.  
These parameter values correspond to the Amid-I excitations in 
$\alpha$-helices \cite{Chirgadze, KuprKudr, Sct1, ASc}. 

In our numerical studies we have also followed the conventions of Scott 
\cite{Sct1} and so, like him, we have used units in which the energy is 
measured in units of $\hbar\nu_a$, time in units of $\nu_a^{-1}$ and length 
in units of $10^{-11}$m. In this case the dimensionless computer 
values of the  parameters are 
$$ J_{comp} = \frac{J}{\hbar \nu_a} = 
0.145,\quad L_{comp} = 0.231, \quad \chi_{comp} = \frac{\chi \times 
10^{-11}m}{\hbar \nu_a} = 0.318,$$ 
\be 
w_{comp} = \frac{w \times 
10^{-22}m^2}{\hbar \nu_a} = 1.825.  
\label{computunit} 
\ee

The results of the numerical simulations are described below.

First, we have started off the simulations taking as the initial 
conditions the function 
\be 
\Psi_{1,1}\,=\,1.  
\ee 
with all other values 
of $\Psi_{j,n}=0$ and putting $u_{j,n}=0.$ We have then added an 
extra absorptive term into the equation for $u_{j,n}$ and performed 
the simulations until we have reached a stationary state solution.  
The obtained solution described a well defined solitonic state.  Its 
energy was around -0.55067. In Fig. 2 we present the plots of the 
$\Psi$ and $u$ fields.

%WOJTEK' PLOT OF PSI and U FIELDS 
\begin{figure}[htbp]
\unitlength1cm \hfil
\begin{picture}(16,6.5)
 \put(4.5,-0.5){\makebox(0,0){2.a}}
 \put(12.5,-0.5){\makebox(0,0){2.b}}
 \epsfxsize=6cm \put(0,6){\rotatebox{-90}{\epsffile{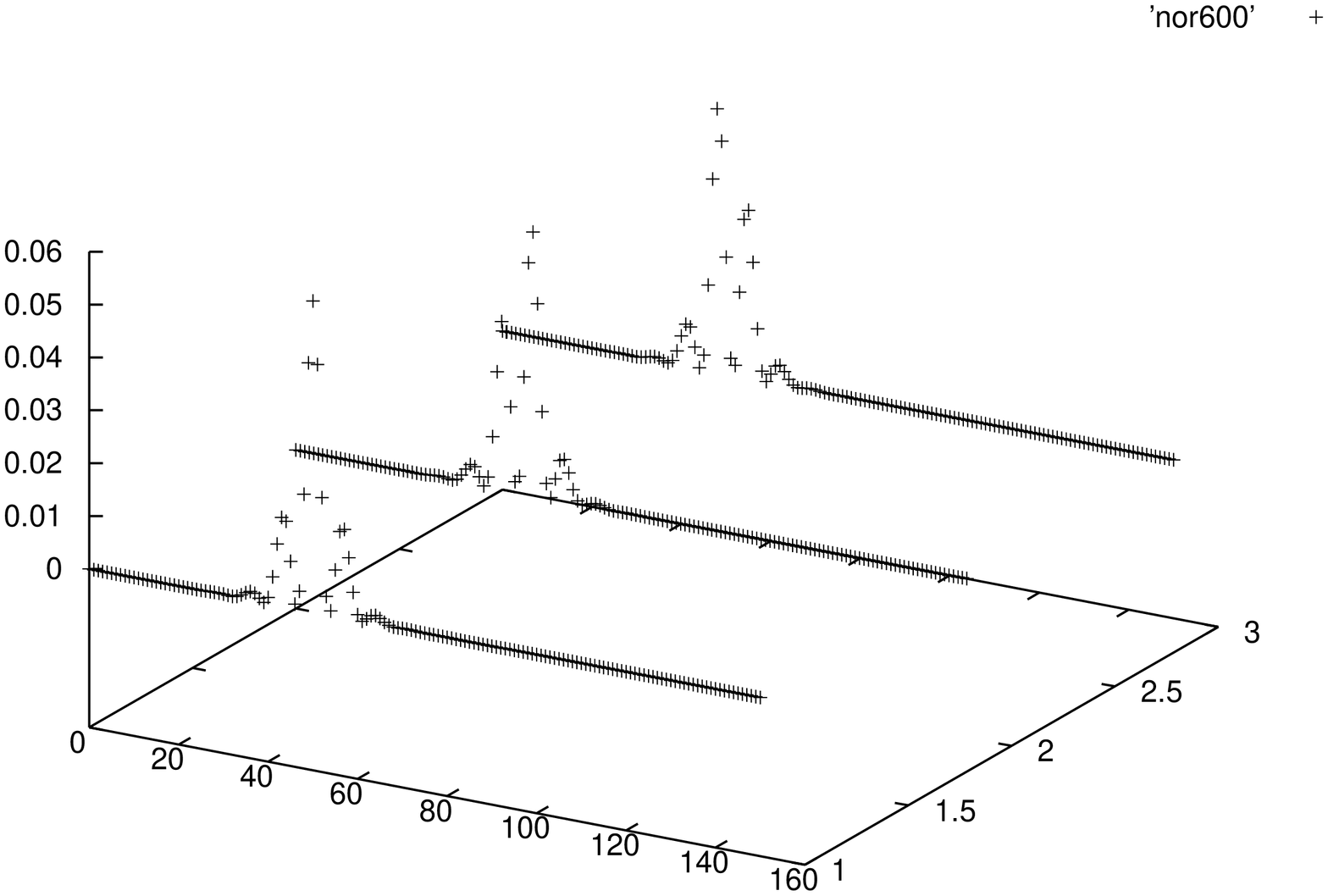}}}
 \epsfxsize=6cm \put(8,6){\rotatebox{-90}{\epsffile{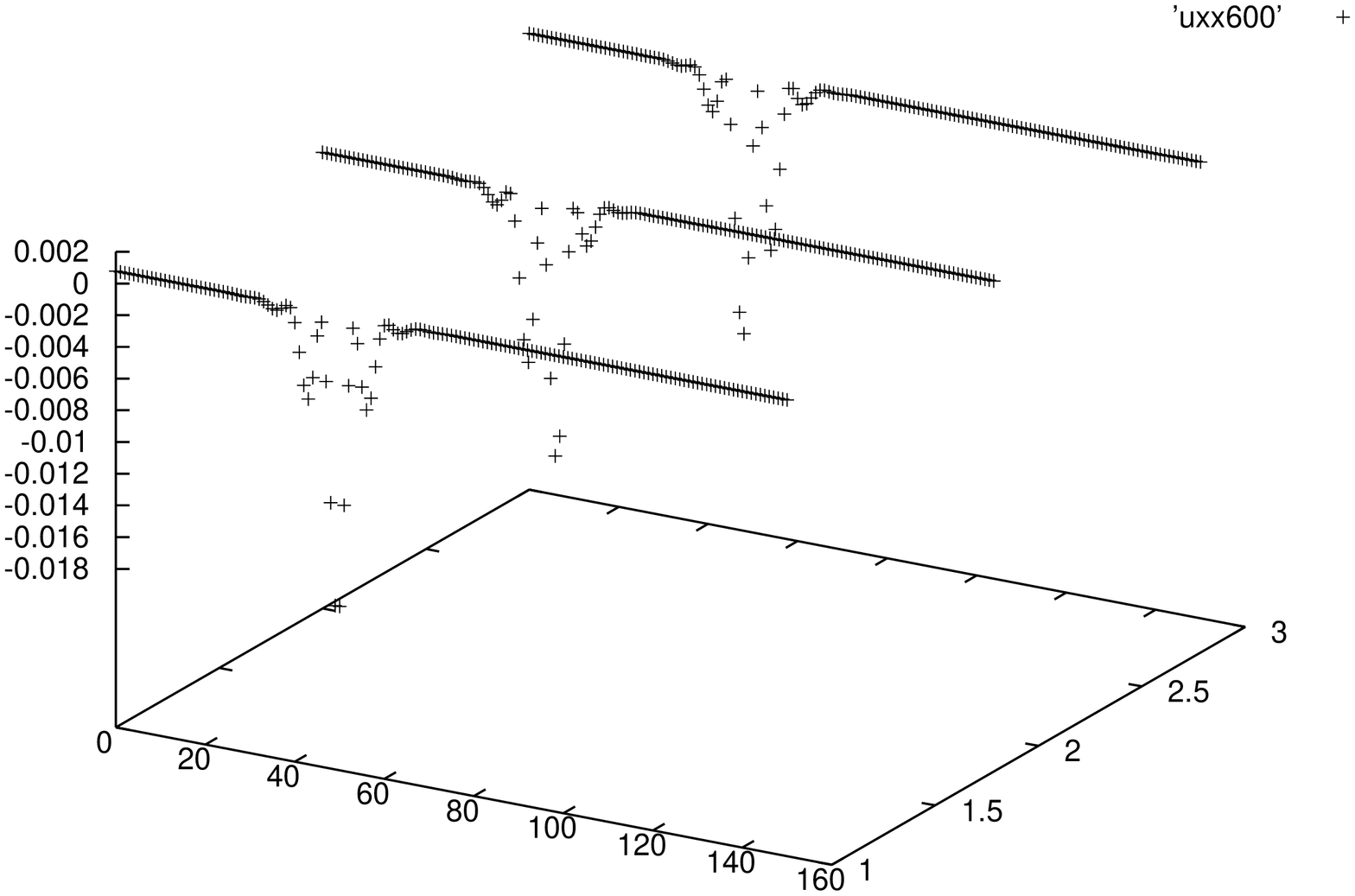}}}
\end{picture}
\vskip 5mm
\caption{Stationary excitation function $\Psi$ (a) and the derivative 
of the PGs displacement {\it i.e.} $u_{j+1,n}-u_{j,n}$ (b) for the 
alpha-helix.} 
\end{figure}

We see that this self-trapped state has an inner structure. While 
the total (summed over all three spines) distribution of the excitation 
has a single-hump pattern, the distributions in individual spines are 
modulated in the manner of solutions (\ref{dpr3}). The same feature can 
also be seen in Fig.6 of \cite{Sct1}. Thus, we can conclude that our 
numerical solution, as well as the solution discussed in \cite{Sct1}, 
describes the lowest energy of the hybrid solitons. This view is 
confirmed also by the numerical estimate of the soliton energy 
(\ref{erest2}). Thus taking our numerical values (\ref{computunit}), 
we get ${\cal{E}}_2(0) - E_0 = -0.55062$ in units of $\hbar \nu_a$ 
which coincides with the value determined in our numerical 
simulations.

Having found stationary solutions, we then changed the functions as 
follows
\be
\Psi_{j,n}\,\rightarrow \Psi_{j}\,e\sp{i n \Delta k }\quad 
\frac{du_{j,n}}{dt}\,\rightarrow\,\frac{d u_{j,n}}{dt}\,+
\,(u_{j,n}-u_{j,n-1}) \sin(\Delta k)
\ee
leaving $u_{j,n}$ unchanged. This had the effect of giving a small speed to
the soliton, and the distortion of the chain.

We then performed the simulation over a short period of time. During this time
the soliton has been moving and the small disturbance introduced by the
nonperfect transfer of momentum to the system has spread itself over the 
lattice.

We then repeated the whole process several times thus slowly increasing
the total $k$ (in practice we put 
$\Delta k$=0.1). After every step we evaluated the resultant speed of the 
soliton. 
Of course, the whole process suffered by the introduced disturbances;
thus gradually it has become more and more difficult to determine
this speed. However, we have found that each addition of momentum increased
this speed by a decreasing amount suggesting that there is a maximum speed
that the soliton can attain. 

\begin{figure}[htbp]
\unitlength1cm \hfil
\begin{picture}(8,8)
 \epsfxsize=8cm \epsffile{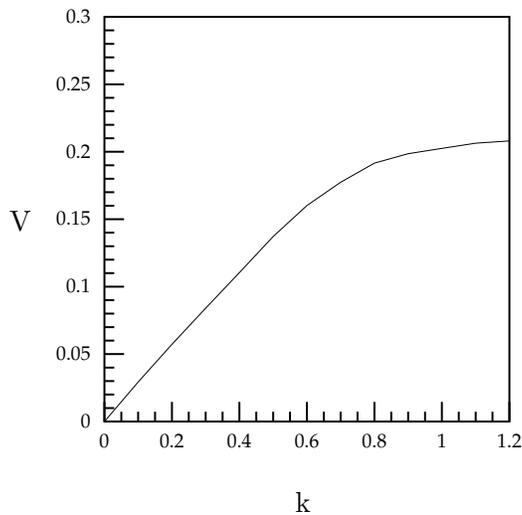}
 \put(-3.5,0.75){k}
 \put(-7.3,4.5){V}
\end{picture}
\caption{Speed of the hybrid soliton as a function of $k$. }
\end{figure}

In Fig. 3 we present a plot of the resultant speed as a 
function of the total $k$ (i.e., the sum of all  $\Delta k$). We note 
that the maximum speed appears to be around 0.21. To check that this 
limit is not an artifact of our procedures, we have performed further 
simulations in which we modified the steps  $\Delta k$ or eliminated the
modification of $\frac{u_{ij}}{dt}$. We have also performed some 
simulations with absorption: the configurations were alternatively 
boosted and then evolved in time but with a small absorption 
parameter added to the equations. These extra terms absorbed some of 
the ripples while the boosts were effectively accelerating the 
solitons.  All these procedures produced similar results and we have 
never managed to get the solitons move faster than with $v\sim 0.21$. 
The absorptions did decrease the deformations of the $\alpha$-helix but they 
did also reduce the velocity of the soliton; hence we do believe that 
the solitons cannot have larger velocity and that this maximum speed 
is determined by the maximum allowed group velocity of the excitons. 

In Fig. 4 we present the plot of the solutions of 
(\ref{carvel}) for $E_{\pm}(k)$ with the values of $J$ and $L$ given 
in (\ref{computunit}) (we recall that $v=V/\nu_a$). From Fig. 4 
we see that, indeed, the composite soliton cannot have its velocity 
larger than the maximum group velocity for one of its two components,  
and for our parameters this velocity is about 0.21. At wavenumber 
$k_{cr}$, which corresponds to the maximum group velocity, 
$d^2E_{\mu}(k)/dk^2 = 0$ and at $k \geq k_{cr}$ the balance between
the nonlinearity and the dispersion breakes down for one of the  
components and this leads to the  decay of the soliton.

%HERE WE PUT THE PICTURE (SECOND) MADE BY BERNARD (fig.4)
\begin{figure}[htbp]
\unitlength1cm \hfil
\begin{picture}(8,8)
 \epsfxsize=8cm \epsffile{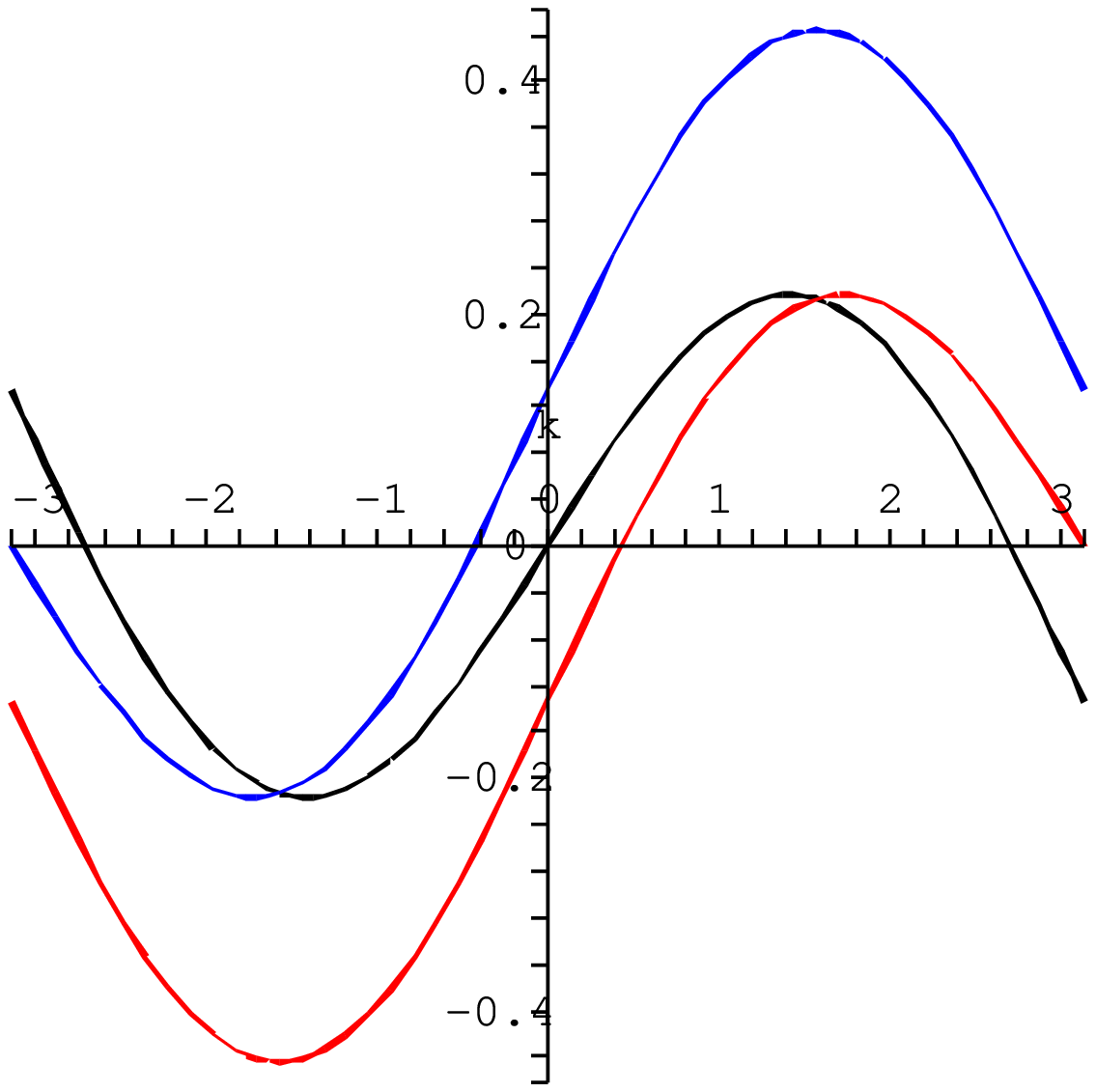}
 \put(-1.3,3){$E$}
 \put(-1.5,6.5){$E_+$}
 \put(-1.5,5){$E_-$}
\end{picture}
\caption{The excitation velocity (\ref{carvel}) for $J=7.8cm^{-1}$ 
and $L=12.4cm$.} 
\end{figure}

The complex (modulated many-hump) and composite (three-spine 
distributed) structure of the soliton manifests itself distinctly when 
the soliton is moving  and the interspine oscillations take place 
(\ref{probch}). This is seen very clearly in the oscillations of 
the probability distribution amplitude for each spine which is shown in 
Fig. 5. This phenomenon
was already noted by Scott in \cite{Sct1}.  According to 
(\ref{period}), the frequency of these oscillations is determined by 
$k_d$ and by the soliton velocity.
It follows from  (\ref{k*}) that the bottom of the band
is attained at $k_d=0.42$.

%HERE THE PICTURE OF FREQUENCY (OR PERIOD) DEPENDenCE ON VELOCITY - Fig.5
\begin{figure}[htbp]
\unitlength1cm \hfil
\begin{picture}(8,8)
 \epsfxsize=8cm \epsffile{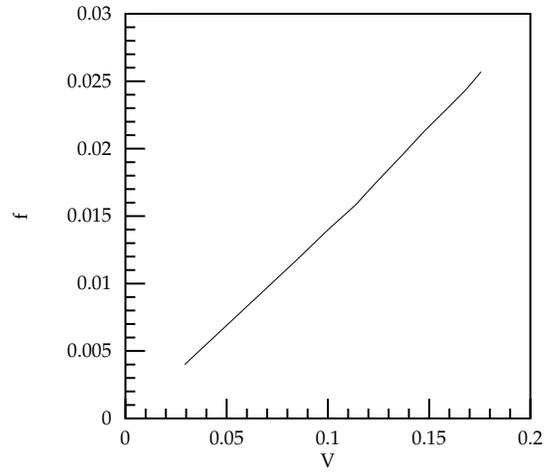}
\end{picture}
\caption{Oscillation frequency of the hybrid soliton as a function of its 
speed $V$.}
\end{figure}

We have also looked at the other two solitons and tried to make them move.
As has already been mentioned above, an arbitrary initial configuration, in 
the case
of a helical structure always leads to the solution of the
lowest energy. But when we take as an initial condition $\Psi_{j,n}$ in
the form (\ref{siterepmu}) at $V=0$,
we have obtained, as a result of the  calculations, stationary solutions and
these solutions  were very close to those derived in the continuum
approximation. The energies of these stationary solutions were
$+0.171$ for $\mu = 0$ and $-0.54972$ for $\mu = \pm 2\pi /3$ which, again,
coincide with the values ${\cal{E}}_{\mu}(0) - E_0$ estimated from
(\ref{erest0}) and (\ref{erest1}). For these states the probability
distribution in individual spines $\cmod \Psi_{j,n}\ $ shows a one-hump
pattern without any modulation (see fig 6). This differentiates these states 
from the lowest
energy composite soliton.

%HERE WE PUT IN MY FIGURE OF OTHER SOLITONS (fig6)
\begin{figure}[htbp]
\unitlength1cm \hfil
\begin{picture}(16,6.5)
 \put(4.5,-0.5){\makebox(0,0){6.a}}
 \put(12.5,-0.5){\makebox(0,0){6.b}}
 \epsfxsize=6cm \put(0,6){\rotatebox{-90}{\epsffile{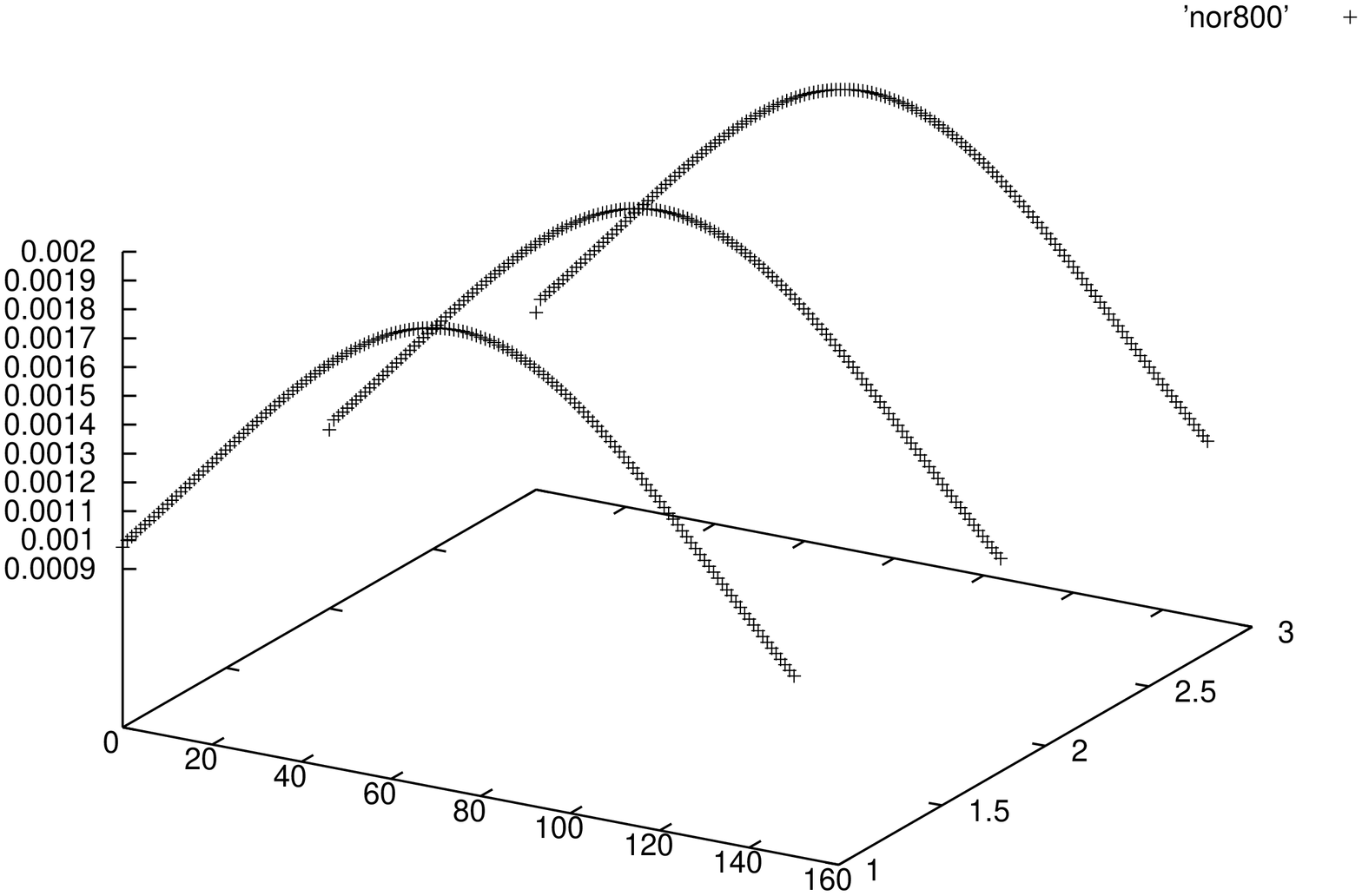}}}
 \epsfxsize=6cm \put(8,6){\rotatebox{-90}{\epsffile{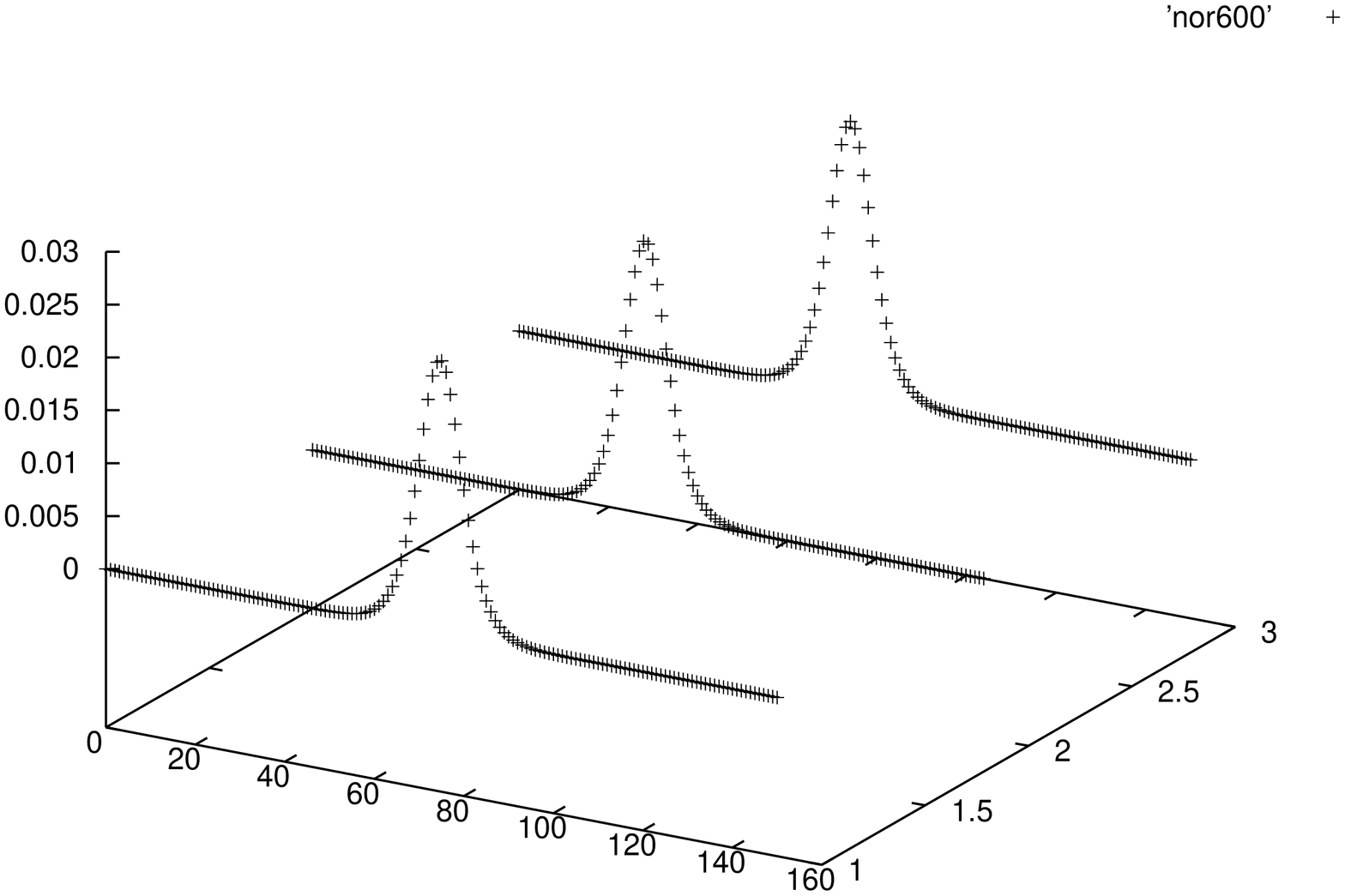}}}
\end{picture}
\vskip 5mm
\caption{Electron fields for the other solitons - (71) with $V=0$.}
\end{figure}

We have tried to make these states move.
Unfortunately, the perturbations introduced by the discreteness of
the lattice and by the inexactness of our procedure led to their
instability.  This showed itself in the system evolving into the
lowest energy (totally symmetric) soliton.

\section{Conditions for the applicability of the adiabatic approximation}

Having found the three types of solutions described in the previous sections a 
question then arises about
the conditions of the applicability of the adiabatic approximation
in such a three-spine model.

The Hamiltonian (\ref{Htot}) that describes the states of 
quasiparticles which interact with phonons, does not have an 
exact solution. The adiabatic approximation describes the 
soliton-like states of large polarons when the autolocalization, 
within the region of several lattice sites, takes place. This is  
one of the three possible approximations which allow us to represent 
the Hamiltonian (\ref{Htot}) as a sum of two terms: the main part, 
$H_0$, and the term, $H_1$, which can be considered as a small 
correction, and, therefore, for which the perturbation theory can be 
developed.  The other two approximations correspond to the almost 
free quasiparticles and to small polarons. The realization of one or 
another of these three regimes depends on the relation between the
parameters of the system. In  general  the problem can be 
investigated in the framework of the variational approach  
\cite{BELmP, BELm, BE}.   The ground state diagram for 
a simple chain with one exciton band and one phonon mode 
was presented in \cite{BELmP}. This diagram showed the range of 
values of the dimensionless coupling constant and of the 
nonadiabaticity parameter (relation $\hbar \nu_a /(2J)$) for which 
one or the other regime was realized.

As it has often been mentioned, various properties of the Davydov solitons in
$\alpha$-helical proteins have been analysed using a
single chain model. Although such a model gives good
qualitative and sometimes also good quantitative \cite{Sct1} 
properties of Davydov solitons, the ground state diagram \cite{BELmP,
BE} shows that the parameters of the $\alpha$-helix  applied to 
a one-chain model, correspond to the state far from the region where 
the soliton ground states are realized.  This is one of the reasons why the  
estimates by H. Bolterauer 
\cite{nato} and J.W.  Schweitzer and J.P.  Cottingham \cite{nato} of the 
Davydov soliton life-time, obtained within a different 
approach but still based on the one-chain model, give very small 
values.

Here we return to this problem and we assess the conditions of the 
aplicability of the adiabatic 
approximation for the $\alpha$-helix basing our discussion, for simplicity,
 on the solutions describing solitons at rest.  
Applying a unitary transformation, the Hamiltonian (\ref{Htot}) 
takes the form $H = H_{ad} + H_{na}$ where $H_{ad}$ 
is diagonal in the new represantation and describes the adiabatic states 
of the exciton (electron)-phonon system. The term $H_{na}$ is 
an operator of nonadiabaticity which describes phonon-induced 
transitions between adiabatic states. Such a transformation was 
used in \cite{Er1, ErGaidVakh} and, based on it, a method of 
partial diagonalization was further developed in \cite{BrizhEr, 
Schweitz, Clogst}. 
 
The partial diagonalization shows  
clearly that the state-vector (\ref{adappr}) is an eigenstate of 
$H_{ad}$ with the eigenenergy ${\cal{E}} = \hbar \Omega + W$ (here 
$W$ is the energy of lattice deformation) provided that the functions 
$\psi _{\mu,k}$ and $\beta_{\nu,q}$ are stationary solutions of 
equations (\ref{eqpsi}) and (\ref{eqbeta}), i.e. $ 
H_{ad}|\psi_0\rangle = {\cal{E}}_S |\psi_0\rangle $. The virtual 
excited adiabatic states, for a given chain deformation (\ref{steqQ}) 
can be found from the linear equation (\ref{eqpsi}).

 If $H_{na}$ is small it is 
possible to construct the perturbation series $$|\psi\rangle 
=|\psi_0\rangle + |\psi_1\rangle + ... $$ where $|\psi_0\rangle = 
|s\rangle $ is the wavevector (\ref{adappr}) of the soliton state in  
the zero order of the adiabatic approximation and $|\psi_i\rangle$ is the  
$i$-th correction due to $H_{na}$. According to the general theory of 
perturbations the first correction is  given by
\be 
|\psi_1\rangle = - 
\frac{Q}{a} H_{na} |\psi_0\rangle 
\label{frstcorr} 
\ee 
where we have defined
$$ \frac{Q}{a} = Q \frac{1}{H_{ad}-{\cal E}_s} Q $$ and 
$$ Q = 1 - |\psi_0\rangle \langle \psi_0| = \sum_{\alpha \neq s} 
|\alpha\rangle \langle \alpha|.$$ 

 Note that here $|\alpha\rangle$ ennumerates  
all adiabatic terms of $H_{ad}$, $H_{ad}|\alpha\rangle = {\cal 
E}_{\alpha}|\alpha\rangle$.  For the convergence of the perturbation 
series  $|\psi_i\rangle$ should be  proportional 
to $\lambda ^{i}$ with $\lambda$ being a small parameter.

 The square of the norm of 
vector $|\psi\rangle$ is 
$$\langle\psi|\psi\rangle = 
\langle\psi_0|\psi_0\rangle + \langle\psi_1|\psi_1\rangle + ... = 1 + 
O(\lambda ^2). $$ 
Therefore, the applicability of the adiabatic 
approximation is guaranteed  provided that
\be 
\langle\psi_1|\psi_1\rangle \equiv  \lambda ^2 \ll 1 .  
\label{adcond} 
\ee 

Taking into account (\ref{frstcorr}), we can calculate 
\be 
\langle\psi_1|\psi_1\rangle = 
\langle\psi_0|H_{na}\frac{Q}{a^2}H_{na}|\psi_0\rangle = \sum_{\alpha}
\langle\psi_0|H_{na}|\alpha\rangle
\langle\alpha|\frac{Q}{a^2}|\alpha\rangle
\langle\alpha|H_{na}|\psi_0\rangle \nonumber\ee
\be
= \frac{\sigma}{\Delta ^2}
\sum_{\alpha} f_{\alpha} \langle\psi_0|H_{na}|\alpha\rangle
\langle\alpha|H_{na}|\psi_0\rangle.
\label{normpsi1}
\ee
Here we have taken into account the fact that  the 
operator $\frac{Q}{a^2}$ is diagonal and we have defined  
\be 
f_{\alpha} = 
\frac{1}{\sigma}\langle\alpha|\frac{\Delta ^2 Q}{a^2}|\alpha\rangle
\label{f_alph}
\ee
with $\Delta$ being the energy gap between the solitonic energy level and 
the lowest excited one.  In (\ref{f_alph})  
\be
\sigma = \sum_{\alpha \neq s} \langle\alpha|\frac{\Delta ^2
Q}{a^2}|\alpha\rangle
\label{sigm}
\ee
so that $\sum_{\alpha \neq s} f_{\alpha} = 1 $. In (\ref{normpsi1}) the
summation does not include $\alpha = s$ because the diagonal matrix elements
of the nonadiabaticity operator vanish. Next we observe that
\be
\sum_{\alpha} f_{\alpha} \langle\psi_0|H_{na}|\alpha\rangle
\langle\alpha|H_{na}|\psi_0\rangle \leq
\sum_{\alpha} \langle\psi_0|H_{na}|\alpha\rangle
\langle\alpha|H_{na}|\psi_0\rangle  =
\langle\psi_0|H_{na}^2|\psi_0\rangle  .
\label{sumtoevr}
\ee
Moreover, it is easy to see that
\be
 \langle\psi_0|H_{na}^2|\psi_0\rangle  =
\langle\psi_0|H^2|\psi_0\rangle - \langle\psi_0|H|\psi_0\rangle ^2 =
\Delta E^2 .
\label{Hna2}
\ee

Thus we can derive some  estimates without the partial 
diagonalization of Hamiltonian (\ref{Htot}). In particular, we can 
calculate $\Delta E^2$ using the soliton wavefunction (\ref{adappr}) 
in the zero order adiabatic approximation.  This way we can estimate 
the soliton life-time in one chain and we get the same result as that 
obtained by J.W.  Schweitzer and J.P. Cottingham \cite{nato} 
 who calculated $\Delta E^2$ performing the partial diagonalization, 
 and by Bolterauer \cite{nato} who calculated $\Delta E^2$. 

So, the condition of the applicability can be writen as
\be
\langle\psi_1|\psi_1\rangle \leq  \frac{\sigma \Delta E^2}{\Delta ^2}
\le 1 .
\label{cond}
\ee
Calculation of $\Delta E^2$ gives us
\be
\Delta E^2 = \frac{1}{2}\left( A - \sum_{\nu,q} \hbar \omega_q^3
|Q_{\nu}(q)|^2 \right),
\label{DE2}
\ee
where
\be
A = \sum_{\nu,q} \frac{4 \hbar \chi ^2 \sin ^2{q}}{3MN\omega_q} =
\frac{16\hbar \nu_a \chi ^2}{3 \pi w}.
\label{cnstA}
\ee
Taking into account (\ref{steqQ}) we can rewrite (\ref{DE2}) in the
form
$$\Delta E^2 = \sum_{\nu,q}\frac{2\hbar \chi ^2 \sin ^2{q}}{3MN\omega
_q}\left(1 - |\sum_{\mu,k}\psi_{\mu,k} \psi_{\mu+\nu,k+q}|^2 \right)$$
which corresponds to Bolterauer's \cite{nato} expression after the 
transformation to the site representation.

For symmetrical solitons $Q_{+}(q)=Q_{-}(q) = 0$  and
\be
Q_0(q)=\frac{2i\chi \sin{q}}{\omega_q \sqrt{3MN}} 
\sum_{k}\psi_{\mu}^*(k+q) \psi_{\mu}(k) = 
\frac{2i\chi \sin{q}}{\omega_q \sqrt{3MN}} \int_{-N/2}^{N/2} e^{iqx} 
|\varphi _{\mu}(x)|^2 dx,
\ee
where 
\be
\int_{-N/2}^{N/2} e^{iqx} |\varphi _{\mu}(x)|^2 dx = 
\frac{\pi q}{2 \kappa_{\mu}\sinh{\frac{\pi q}{2\kappa_{\mu}}}}.
\ee
Therefore,
\be
\Delta E^2_1 = \frac{8\hbar \nu_a \chi ^2}{3\pi w}(1 - \frac{1.8}{\pi 
^2}\kappa _{\mu}^2 ).
\ee

For the hybrid soliton we have
\be
Q_0(q)=\frac{2i\chi \sin{q}}{\omega_q \sqrt{3MN}} \left(
\int_{-N/2}^{N/2} e^{iqx} |\varphi _{+}(x)|^2 dx +
\int_{-N/2}^{N/2} e^{iqx} |\varphi _{-}(x)|^2 dx \right)
\ee
and 
\be
Q_{+}(q)=Q_{-}^{*}(-q) =
\int_{-N/2}^{N/2} e^{i(q+2k_d)x} \varphi _{+}^{*}(x)\varphi 
_{-}(x))dx, 
\ee
\be
\Delta E^2_2 = \frac{8\hbar \nu_a \chi ^2}{3\pi w}(1 - \frac{7.2}{\pi 
^2}\kappa _{2}^2 - \frac{\pi}{6}\kappa_{2}\sin{k_d}\cos^2{k_d} ).
\ee

In the one-dimensional case, the soliton
level (121) is a single bound level in the lattice deformation
potential. Excited adiabatic states belong to the quasi-continuum 
spectrum with eigenenergy $\Lambda(k) = \frac{\hbar ^2 k^2}{2m_{\mu}}$ 
 which is separated from the soliton level by a gap $\Delta = 
\frac {\hbar ^2 \kappa_{\mu}^2}{2m_{\mu}}$. Therefore, we can 
estimate $\sigma$ (\ref{sigm}) as
\be
\sigma = \frac{n}{N} \sum_{k} \frac {\kappa_{\mu}^4}{(\kappa ^2 + 
k^2)^2} = \frac{n \kappa_{\mu}}{4},
\ee
where $n=1$ for the totally symmetric soliton, and $n=2$ for the 
hybrid soliton since, in this case, there are two degenerate bands. 

Finally, we have the conditions for the realization of the soliton-like 
states: 
\be
\lambda_0 = \sqrt{\frac{2C_0}{\pi}} \frac{w \sqrt{\hbar \nu_a 
(9J-L)}}{\chi ^2} < 1
\label{cond1}
\ee
for the total symmetric soliton, and
\be
\lambda_1 = \sqrt{\frac{2C_1}{\pi}} 
\left(\frac{2}{3}\right)^{\frac{3}{2}} \frac{w \sqrt{\hbar \nu_a 
(18J+L)}}{\chi ^2} < 1
\label{cond2}
\ee
for the composite soliton, respectively. Here 
$$ C_0 = 1 - \frac{1.8}{\pi ^2}\kappa _0^2  $$ 
and 
$$ C_1 = 1 - \frac{\pi}{6} k_d \kappa_2 - \frac{7.2}{\pi ^2}\kappa 
_2^2 $$ 
where we have assumed that $k_d \ll \pi$ and $\kappa _{\mu} < 1$. 

Note that the condition (\ref{cond1}) coincides with the condition 
which can be obtained in the partial diagonalization scheme for the one 
chain model \cite{Schweitz, BrizhEr}. This condition indicates that 
solitons can exist in soft enough chains and at a strong enough 
exciton (electron)-phonon coupling they are stable against quantum 
fluctuations. The relation (\ref{cond1}) is the inverse of the condition 
for the weak coupling regime. 

The numerical values of the parameters for the $\alpha $-helix are:  
$J=1.55 \cdot 10^{-22}$ Joule, $L=2.46 \cdot 10^{-22}$ Joule, $\chi 
=35\ -\ 62 \ \cdot 10^{-12}$ N, $w_H=13\ -\ 19.5$ N/m. We can take  
$1/\nu _a=10^{-13}$ s.  For these parameters we get $\lambda_0 =2.3\ 
-\ 11$ for the total symmetric soliton, which corresponds to the 
one-band model. Therefore, in this case, the adiabatic approximation 
is not valid, and, consequently, the soliton is destroyed by quantum 
fluctuations. The corresponding estimates for the composite soliton 
give the value $\lambda _1 \approx 1$. For instance, for $\chi =62 \ 
\cdot 10^{-12}$ N and $w_H=14.6$ N/m we get $\lambda _2=0.87$, and,  
therefore, the perturbation series converges. It is worth adding here 
that the larger values of the coupling and the condition that the chains are 
softer strengthen the condition for this type of soliton solution to exist.

\section{Conclusions}

As we have mentioned above, the main aims of this paper were the study 
of the soliton states in $\alpha $-helical proteins taking into account  
their helicity structure and the understanding of the origin 
of the inter-spine oscillations observed numerically by Scott in 
\cite{Sct1}. The soliton states 
in $\alpha$-helical proteins are described by a system of 
nonlinear  equations (\ref{-1}-\ref{beta}). In our study we have restricted 
the Hamiltonian of amide excitations to two main terms, namely, those that 
describe the intra- and 
inter- spine interactions, while Scott considered ten additional 
terms of long-range resonance interactions.  Our results broadly 
reproduce the results of Scott. However, there are also some 
differences, which we summarize below.

The velocity of the soliton propagation in the numerical
calculations carried out by Scott in \cite{Sct1}, was reported as
$V=\frac{3}{8}\nu _a$, while our results give the maximum value 
$V=0.21 \nu _a$. This is due to the fact that in \cite{Sct1} 
 further terms of the resonance interaction of Amid-I vibrations 
were included, which increase the width of the exciton bands and, 
therefore, increase the exciton group velocity. The additional 
 terms in the Hamiltonian also change  the corresponding value of 
$k_d$, but, probably, this change is less significant than the change 
of the maximum group velocity. Nevertheless, our formula (\ref{period})  
of the period of oscillations for the values  
$\nu _a=10^{13}s^{-1}$ and  $k_d=0.42$ for the $\alpha $-helix at 
$V=\frac{3}{8}\nu _a$, gives $T=1.995 \cdot 10^{-12} s$, which
practically coincides with the value obtained by Scott, 
$T_{comp}=2 \cdot 10^{-12} s$.  

Our analytical study and the numerical simulations elucidate the 
conditions for the 
existence of various types of soliton solutions: single-band and 
mixed two-band solitons. The entangled two-band (hybrid) solitons 
break spontaneously the translational and rotational symmetries, and 
possess the lowest energy.  Single-band solutions break only the 
translational symmetry and preserve the rotational symmetry.  
Single-band solitons turn out to be dynamically unstable: once 
initially formed, they decay rapidly while propagating. There are two 
main reasons for this, which arise from the helical structure of the 
system, namely, the absence of the forbidden gap in the energy 
spectrum (see Fig. 1) and the Umklamp processes. The absence of the 
energy gap allows the transition to the lowest energy state via the 
interactions with low-energy phonons. The helical symmetry leads to 
the relation
$\psi _\mu (k \pm 2\pi )=\psi _{\mu \pm 1}(k) $, i.e.,   
the mixing of single-band states takes place, and, as a result, 
the single-band solutions decay. This is the reason why given any 
initial condition the excitation localises into the state which 
corresponds to the lowest energy, i.e., to the entangled soliton. In 
particular, this was the case observed in \cite{Sct1}. We have
managed to observe such solitons due to the very special choices of the 
initial excitations  which were very close to the expression for the stationary
single-band soliton at rest.  

It is also worth comparing this type of solution in a helical  
system with those in a three-spine model without helical 
symmetry. In the latter case there is a forbidden gap in the energy 
spectrum between the two degenerate bands and the third band. As a 
result, the initial excitation with the energy above the forbidden 
gap, is self-trapped in a single-band soliton state. The totally 
symmetric soliton predicted analytically in \cite{DES,Er-Serg}  
was observed numerically in \cite{Hyman,Hen}.  Such a soliton in a 
chain without helical symmetry can be destroyed only if a large amount of 
energy is supplied to the system.  Therefore, these single-band 
localised solutions are  much more stable dynamically than single-band 
solitons in chains with helical structures.  This constitutes a 
qualitative difference between the three-chain system with an 
helical symmetry and the one without it. 

The important question about the existence of Davydov solitons in 
$\alpha $-helical proteins remains open. Unlike the case of 
conducting polymers, for which there is reliable 
experimental evidence for the soliton (large polaron and bipolaron) 
existence, such data are absent for polypeptides.  The answer to this 
question is related to the applicability of the 
adiabatic approximation, which is determined by the numerical values 
of the parameters of a given system. Solitons 
can exist in protein macromolecules provided their parameters satisfy 
the condition of the adiabatic approximation.  Note, e.g., that the 
spring constant for the hydrogen bond $w_H$ was determined in \cite{KuprKudr}
 to be $21$  N/m.  Scott \cite{Sct1}, who takes into account that 
the hydrogen bonds in the $\alpha$-helix are $22^o$ oriented, uses the 
value $19.5$ N/m. But as it has been shown above, the effective value is 
$w=\gamma^2 w_H$ where $\gamma$ is determined in (\ref{gamma}). For 
the parameters of the $\alpha$-helix $a = 5.4 \AA$ and $r = 1.7 \AA$ 
we get $\gamma = 0.9$ and, therefore, $w = 17.05$ N/m. Thus, the 
geometrical factor helps to satisfy the condition for the exsistence of a 
soliton.

As we have seen above, the 
generally accepted parameters for Amid-I excitation do not favour the 
existence of single-band solitons. On the other hand, they are proper for 
the existence of the entangled soliton states, although the 
nonadiabatic corrections are also important and ought to be taken 
into account. Thus, the one-chain 
model can give good qualitative results, but conclusions concerning 
the existence and stability of soliton states, based  on numerical 
calculations within such an oversimplified model, may not always be correct.
Of course, our estimates are relatively rough, and the 
method of partial diagonalization of the Hamiltonian would provide 
better results. Its generalization to systems involving 
three-chain macromolecules can face the problem of the applicability of the
long-wave approximation. In such cases the partial 
diagonalization method developed for discrete models by Clogston et 
al. \cite{Clogst} may turn out to be useful. Moreover, the variational methods can give better results (see, e.g., 
\cite{Brown,BELm,BELmP, BE}) for the 
crossover states when the perturbation scheme parameter is not very 
small.

\section{Acknowledgement} 
This work has been supported by a Royal Society grant.

{}

\end{document}